\documentclass[
twocolumn,
english,
aps,
pra,
reprint,
nofootinbib,
superscriptaddress,
floatfix
]{revtex4-2}

\pdfoutput=1

\usepackage{graphicx}
\usepackage{dcolumn}
\usepackage{bm}
\usepackage{color}
\usepackage{xspace}
\usepackage{babel}
\usepackage{subfigure}
\usepackage{accents}
\usepackage{verbatim}
\usepackage{placeins}
\usepackage{float}
\usepackage{afterpage}
\usepackage{hhline}
\usepackage{natbib}

\usepackage{hyperref}

\hypersetup{
    colorlinks,
    linkcolor={black},
    citecolor={black},
    urlcolor={blue},
}

\makeatletter
\@ifundefined{textcolor}{}
{%
 \definecolor{BLACK}{gray}{0}
 \definecolor{WHITE}{gray}{1}
 \definecolor{RED}{rgb}{1,0,0}
 \definecolor{GREEN}{rgb}{0,1,0}
 \definecolor{BLUE}{rgb}{0,0,1}
 \definecolor{CYAN}{cmyk}{1,0,0,0}
 \definecolor{MAGENTA}{cmyk}{0,1,0,0}
 \definecolor{YELLOW}{cmyk}{0,0,1,0}
 \definecolor{PURPLE}{rgb}{0.7,0,0.7}
 \definecolor{dgreen}{rgb}{0,0.6,0}
}

\def\SrOCH{$\textrm{SrOCH}_3$\xspace}
\def\SrNH{$\textrm{SrNH}_2$\xspace}
\def\SrSH{$\textrm{SrSH}$\xspace}

\def\CaOCH{$\textrm{CaOCH}_3$\xspace}
\def\RaNH{$\textrm{RaNH}_2$\xspace}
\def\YbNH{$\textrm{YbNH}_2$\xspace}
\def\BaNH{$\textrm{BaNH}_2$\xspace}

\def\Xt{$\tilde{X}$}
\def\At{$\tilde{A}$}

\def\Xctv{$\tilde{X} ^2A_1$}
\def\Actv{$\tilde{A} ^2E_{1/2}$}

\def\Xcwv{$\tilde{X} ^2A_1$}
\def\Acwv{$\tilde{A} ^2B_{2}$}

\def\Xcs{$\tilde{X} ^2A'$}
\def\Acs{$\tilde{A} ^2A'$}

\def\ctv{$C_{3v}$\xspace}
\def\cwv{$C_{2v}$\xspace}
\def\cs{$C_s$\xspace}

\newcommand{\X}{\tilde{X}^{2}}
\newcommand{\A}{\tilde{A}^{2}}
\newcommand{\B}{\tilde{B}^{2}}
\newcommand{\C}{\tilde{C}^{2}}
\newcommand{\wn}{$~\rm{cm}^{-1}$}
\newcommand{\s}{^\mathsection}

\usepackage{pslatex} 

\begin{document}

\title{Vibrational Branching Ratios for Laser-Cooling of Nonlinear Strontium-Containing Molecules}

\author{Alexander Frenett}
\email{afrenett@g.harvard.edu}
\affiliation{Harvard-MIT Center for Ultracold Atoms, Cambridge, Massachusetts 02138, USA}
\affiliation{Department of Physics, Harvard University, Cambridge, Massachusetts 02138, USA}

\author{Zack Lasner}
\affiliation{Harvard-MIT Center for Ultracold Atoms, Cambridge, Massachusetts 02138, USA}
\affiliation{Department of Physics, Harvard University, Cambridge, Massachusetts 02138, USA}

\author{Lan Cheng}
\affiliation{Department of Chemistry, The Johns Hopkins University, Baltimore, Maryland 21218, USA}

\author{John M. Doyle}
\affiliation{Harvard-MIT Center for Ultracold Atoms, Cambridge, Massachusetts 02138, USA}
\affiliation{Department of Physics, Harvard University, Cambridge, Massachusetts 02138, USA}

\date{\today}

\begin{abstract}
The vibrational branching ratios from the lowest excited electronic state for \SrOCH, \SrNH, and \SrSH are measured at the $< 0.1\%$ level. Spectra are obtained by driving the \Xt--\At\ transitions and dispersing the fluorescence on a grating spectrometer. We also perform \emph{ab initio} calculations for the energies of vibrational levels relevant for laser cooling, as well as branching ratios to support the interpretations of all molecular spectra. Symmetry group analysis is applied in conjunction with our data to study rotational closure in these molecules. These analyses indicate favorable prospects for laser cooling \SrNH \ and other similar alkaline-earth(-like) amides for future beyond the Standard Model physics searches using polyatomic molecules with long-lived parity doublets.

\end{abstract}

\maketitle

\section{\label{sec:intro} Introduction}

Laser cooled and trapped polyatomic molecules are an attractive platform for precision measurement. Recognition of the long coherence times possible in their electronic ground states has led to proposals for their use in searches for physics beyond the Standard Model (BSM), such as permanent electric dipole moments (EDMs) of fundamental particles~\cite{Hao2020,Kozyryev2021,Isaev2018,Norrgard2019,Kozyryev2017, Isaev2017, Denis2020,Hutzler2020}. The advantage of polyatomic molecules in EDM experiments over simpler laser-coolable diatomic molecules stems from their low-lying closely spaced opposite parity states, known as ``parity doublets"~\cite{Kozyryev2017,Isaev2017,Safronova2018,Hutzler2020,Denis2020}. This structure allows for the full polarization of these species in the lab frame, and hence the full realization of their intrinsic sensitivities to new physics, as well as being powerfully useful in suppressing systematic errors in precision searches for symmetry violation~\cite{Eckel2013,DeMille2015}.

Laser-cooled diatomic molecules (so far BaF, BaH, CaF, CaH, SrF, YbF, and YO \cite{Anderegg2017,Vazquez-Carson2022,Barry2014,Collopy2018,Alauze2021,Zeng2024,mcnally2020optical}) have long-lived ($\tau \gg 10$~s) opposite-parity states arising from the rotation of nuclei around each other (i.e. ``rotational states''). However, there are two major differences between these states and the parity doublet states available in polyatomic molecules. First, the energy separation between rotational states is $\sim$10~GHz, thus requiring an electric field $\mathcal{E} > 10$~kV/cm to achieve near unity polarization.  In contrast, the isoelectronic linear polyatomic species (e.g., CaOH, SrOH, and YbOH) contain partity doublet states arising from vibrational angular momentum around the internuclear axis, and thus can be polarized with fields about 1000 times smaller. Second, these parity doublet states result in a Stark level structure that is fundamentally different than rotational state mixing in diatomic molecules. For example, in the highly polarized limit, the polyatomic molecule structure for vibrationally excited states typically includes two highly and oppositely oriented states, crucial for rejection of systematic errors in EDM experiments, as well as a state with zero orientation~\cite{Kozyryev2017}. 

A disadvantage of using parity doublet states in a vibrational excited state is the shorter spontaneous lifetime compared to diatomic rotational states, $\tau \lesssim 1$~s (e.g. see Ref.~\onlinecite{hallas2023caohtrap}). Symmetric and asymmetric top molecules, which we refer to as nonlinear molecules, feature many of the same desirable properties as vibrationally excited linear polyatomic molecules (low-field polarizability, parity-doublet-like structure, and compatibility with laser cooling~\cite{Kozyryev2016,Augenbraun2020,Hutzler2020}), while having intrinsic lifetimes $\gtrsim 10$ s. The benefits of nonlinear molecules for BSM searches, including the possiblity for very long coherence times, have been pointed out previously~\cite{Kozyryev2016,Augenbraun2020,Hutzler2020}, but key elements of laser cooling them are less explored than for linear counterparts. While 1D laser cooling of a symmetric top species has been demonstrated~\cite{Mitra2020}, and an extension to asymmetric top species has been proposed~\cite{Augenbraun2020} with initial studies already performed in CaNH$_2$ \cite{Burchesky2023}, further work is required to establish a road map to full 3D laser cooling and trapping of such molecules. In addition, the few extant studies of laser cooling low-symmetry molecules have focused predominantly on relatively light molecules. While these species are of interest for quantum information applications, many proposals for BSM physics searches rely on constituent atoms at least as heavy as Sr~\cite{Flambaum1985,DeMille2015,Kozyryev2017,Dzuba2017,Isaev2017,Safronova2018,Hutzler2020,Denis2020,Kozyryev2021}. The effect of mass-dependent perturbations on laser cooling low-symmetry species is not well understood. As such, studying heavier molecules, even in the same symmetry group, provides vital information for precision measurement experiments that is not easy to generalize from existing work. The same or similar molecules are also of interest for quantum information platforms \cite{Wei2011,Wall2013,Wall2015,Yu2019} and the study of low-temperature chemical reactions \cite{Balakrishnan2016,Bohn2017}.

To laser cool any molecule, vibrational loss channels need to be identified to sufficient resolution. Computational techniques are rapidly improving \cite{Li2019,Paul2019,Mengesha2020,Kos20,Zhang2021,Lasner2022,Zhang23,Zhu24}, but predictions of low-probability ($\sim$$10^{-4}$--$10^{-5}$) decay channels have not yet been benchmarked against experimental results for heavy, nonlinear polyatomic molecules where vibronic perturbations are challenging to quantitatively model. As such, \emph{ab initio} methods cannot be solely relied upon to assess the laser coolability of a complex molecule and direct measurements of the vibrational branching ratios (VBRs) must be experimentally recorded.

Though vibrational branching occurs during photon cycling in any molecular species, rotational loss may qualitatively differ in nonlinear species compared to their linear analogs. In linear species, rotationally closed transitions are simple to identify based on parity and angular momentum selection rules, and the ability to drive a single rotational transition for most vibrational repumping levels is ensured~\cite{Bethlem2003, Stuhl2008, DiRosa2004}. Such rotationally closed transitions are not generally possible when laser cooling nonlinear molecules~\cite{Kozyryev2016,Mitra2020,Augenbraun2020}. Rotational leakage channels for laser cooling schemes have not been previously analyzed on the basis of general symmetry-group properties. In particular, whether or not there are symmetry groups which have ``sufficient'' symmetry to maintain rotational closure in reasonable perturbative limits is an open question. 

To address these questions, we measure the VBRs of the lowest excited electronic states in three strontium-containing molecules: \SrOCH, \SrNH, and \SrSH. We measure the vibrational branching to $0.01-0.1\%$, and by doing so investigate the mid-resolution vibrational branching of heavy species in three distinct symmetry classes ($C_{3v}$, $C_{2v}$, and $C_s$). Our measurements also invite an analysis of the rotational structure of the laser cooling transitions proposed for these molecules, and we expand upon ideas presented in Ref.~\onlinecite{Augenbraun2020}. We also point out known perturbations in the Sr-containing species that exacerbate rotational leakage channels compared to the previously studied Ca species.

Through these analyses, this work provides a side-by-side comparison of the technical complexity required to laser cool isoelectronic species as a function of molecular symmetry. The combination of rovibrational analyses indicate that the easiest molecule to laser cool among those studied here is likely \SrNH, the species of \emph{intermediate} symmetry $C_{2v}$. Correspondingly, molecules from the $C_{2v}$ point group such as \SrNH (or heavier analogues such as \BaNH, YbNH$_2$, and RaNH$_2$) appear to be the most preferable species for a next-generation laser-cooled EDM-sensitive nonlinear molecule---having fewer vibrational modes and rotational leakage channels than \SrOCH, but greater symmetry protection against internal perturbations than \SrSH. This highly motivates future work to measure VBRs in \SrNH at the $\sim$$10^{-5}$ level, similar to prior work on SrOH~\cite{Lasner2022}, which should be possible with a few additions to our current methods.

Nevertheless, our work does not show the \textit{impossibility} of laser cooling in any of these species, suggesting that with sufficient motivation laser cooling even $C_s$ (and more easily, $C_{2v}$ or $C_{3v}$) molecules is realizable, in agreement with previous proposals~\cite{Kozyryev2016,Kozyryev2019,Augenbraun2020}.

\section{\label{sec:exp} Method And Apparatus}

\subsection{Experimental}
\begin{figure}
    \centering
    \includegraphics{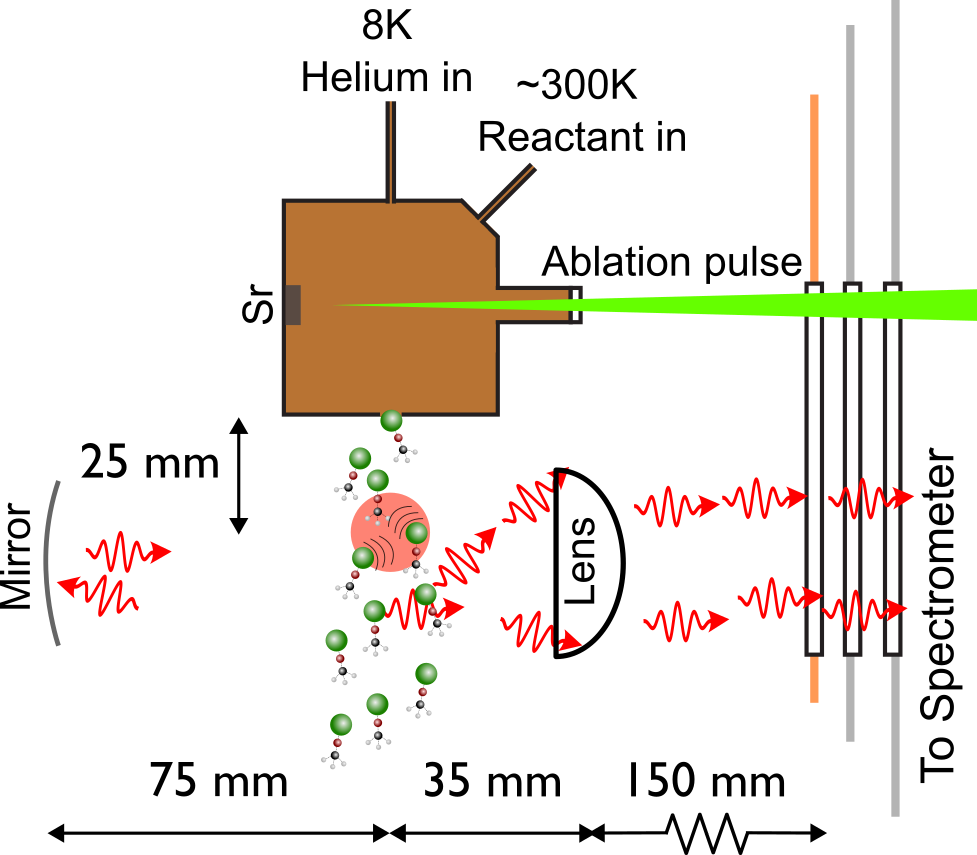}
    \caption{Schematic showing the apparatus as viewed from above. A strontium metal target, located inside a cryogenic copper cell, is ablated by an Nd:YAG laser. The strontium atoms then react with the reactant gas to form the molecule of interest. The molecules are thermalized to the cell temperature by helium buffer gas, and exit the cell rotationally cold. A laser beam excites molecules to the lowest excited electronic state, and the fluorescence from the subsequent decay is collected into a spectrometer. }
    \label{fig:apparatus}
\end{figure}

We measure the VBRs of the three molecular species using the dispersed laser-induced fluorescence method and apparatus used in Ref.~\onlinecite{Lasner2022} for SrOH. The key details are briefly repeated here for clarity. A schematic of the apparatus can be seen in Fig.~\ref{fig:apparatus}.

The molecules are produced in a cryogenic cell ($\sim$8~K) by ablating a strontium target in the presence of both a species-specific reactant gas and a helium buffer gas. Sr atoms released by ablation react to form the molecules of interest, which are then quickly thermalized to the cell temperature by collisions with the buffer gas. Methanol is used as the reactant gas to produce \SrOCH, ammonia for \SrNH, and 1,3-propanedithiol for \SrSH.

About 1'' downstream of the production cell, the molecular beam is intersected by a $\sim$50~mW, $\sim$5~mm diameter beam of excitation light, tuned to the rovibronic transition of interest (discussed in Section \ref{sec:molecules}). The excitation light is generated using a Matisse tuneable Ti:Sapph laser, and is locked to $\sim$5~MHz with a High Finesse WS7 wavemeter. 

The fluorescence from the molecular decays is collimated with a 50~mm in-vacuum lens (focal length 35~mm) and directed into a 0.67~m Czerny-Turner style spectrometer. The overall collection efficiency is limited by the spectrometer’s numerical aperture of 0.11. In the spectrometer, a 2400~line/mm grating disperses a $\sim$40~nm region of the spectrum onto an EMCCD camera. We adjust the grating angle throughout the data taking process to select different subsets of the spectrum and thus cumulatively image over the entire wavelength range of interest for each molecule.

Calibration of the spectrometer and camera system using known frequencies allows us to infer the wavelength and thus vibrational identity of the decays. The relative intensity combined with the spectral response of the spectrometer and camera allows us to obtain the relative probability of each decay.

Off-resonant scatter from the ablation laser, fluorescence from strontium atom decays, and EMCCD signal offsets can all contribute to the fluorescence signal as false molecular decays. We therefore take images of all combinations of ablation laser on/off and excitation light on/off for a given data point. A linear combination of images from these four configurations provides only the spectrum of the light emitted by the target molecules. We collect data using this method long enough to reach an ultimate VBR sensitivity of $0.01-0.1\%$.

\subsection{Computational}

We have performed {\it{ab initio}} calculations for the vibrational levels and branching ratios in SrOCH$_3$, SrNH$_2$, and SrSH. The equilibrium structures and force constants for the electronic ground and first excited states have been calculated at the equation-of-motion electron attachment coupled-cluster singles and doubles (EOMEA-CCSD)~\cite{Stanton93a,Nooijen95} level using closed-shell electronic ground states of the cations as the reference states and adding the open-shell electrons to obtain the targeted electronic states of the neutral molecules. The EOMEA-CCSD calculations treat the ground and excited states of the neutral molecules on the same footing and have been shown to provide accurate results for the vibronic energy levels and branching ratios in metal mono-hydroxides~\cite{Zhang2021,Lasner2022,Zhang23}.

The computed vibrational branching ratios have been obtained by using the computed Franck-Condon factors and transition energies. The calculations of the Franck-Condon factors for the transitions from the vibrational ground state of the electronic first excited states to vibrational ground and excited states of the electronic ground states within the double harmonic approximation have been carried out using the ``FCSQUARE'' module~\cite{Rabidoux16} of the CFOUR program~\cite{CFOURfull,Matthews20a}.

The anharmonic contributions to the vibrational frequencies and wave functions of the electronic ground state have been accounted for using vibrational second-order perturbation theory (VPT2)~\cite{Mills72}.
Fermi resonances have been resolved by means of explicit diagonalization in the space of the quasi-degenerate vibrational modes. 
We have used the GUINEA module~\cite{Kjaergaard08} of the CFOUR program for the VPT2 calculations.
To account for nominally symmetry-forbidden transitions in SrOCH$_3$ and SrNH$_2$, 
we have included the contributions from the ``direct vibronic coupling'' (DVC) mechanism proposed in earlier work~\cite{Zhang2021,Zhang23}, namely, the vibronic coupling between the \Actv\ and $\B A^\prime$ states in SrOCH$_3$ and between the $\A B_2$ and $\C A_1$ states in SrNH$_2$.

We have used the exact two-component (X2C) theory~\cite{Dyall97,Liu09} and the X2C recontracted basis sets to treat relativistic effects.
The spin-free exact two-component theory in its one-electron variant (the SFX2C-1e scheme)~\cite{Dyall01} has been used for the calculations of SrSH and SrNH$_2$. 
Note that the first excited states of SrOCH$_3$ are doubly degenerate in the C$_{3v}$ structure and are subject to Jahn-Teller distortion in the scalar representation. Interestingly, in the spinor representation with spin-orbit coupling included in the molecular orbitals, the lowest excited electronic $^2E_{1/2}$ state of SrOCH$_3$ exhibits a stable $C_{3v}$ structure. This is similar to the observation for RaOCH$_3$~\cite{Zhang23a} and allows the calculations of the vibrational branching ratios using the double harmonic approximation. Therefore, we have used the spinor-based X2C approach with atomic mean-field integrals (the X2CAMF scheme)~\cite{Liu18,Zhang22} for the calculation of SrOCH$_3$. 
The implementation of analytic X2C-EOM-CCSD gradients~\cite{Cheng11b,Zhang23a} in the CFOUR program has greatly facilitated the calculations of equilibrium structures and force constants presented here. 

In the calculations of SrOCH$_3$ and SrSH, we have used the cc-pwCVTZ basis set for Sr \cite{Hill17}, cc-pCVTZ basis set for S~\cite{Woon93,Peterson02}, and cc-pVTZ basis sets for O, C and H~\cite{Dunning89}. 
The treatment of Fermi resonances in SrNH$_2$ is very sensitive to the relative vibrational frequencies of the quasi-degenerate vibrational modes. Aiming at accurate force fields, we have adopted more extensive basis sets for calculations of SrNH$_2$. 
Here we have used the uncontracted ANO-RCC sets~\cite{Faegri01,Roos04a,Roos04} for Sr, N, and H in the calculations of SrNH$_2$, which are of quadruple-zeta quality.

\section{\label{sec:molecules} Rotational Structure and Excitation Transitions}

Since much of the literature on laser cooling molecules is focused on diatomic and linear triatomic species, here we outline the structural considerations of each of the molecules studied in this work. These considerations are relevant to both identifying the excitation transitions used to measure the relevant VBRs, as well as for evaluating the robustness of rotational closure.  

\begin{figure*}
    \centering
    \includegraphics[width = \linewidth]{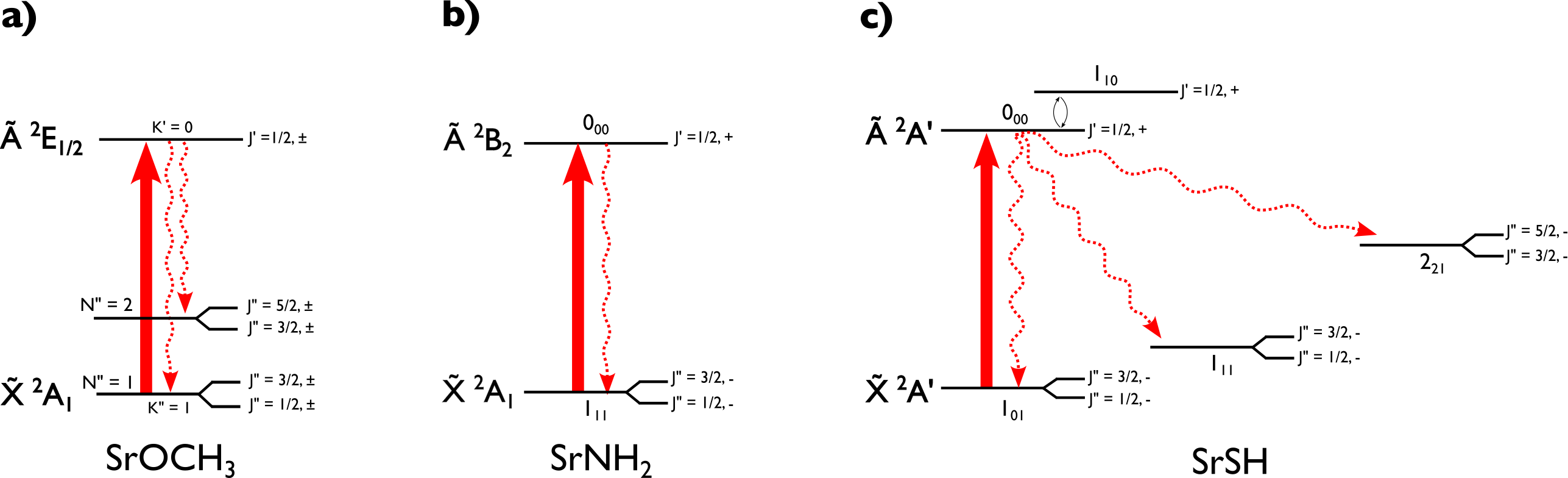}
    \caption{Schematics showing the relevant rotational structure in a) \SrOCH, b) \SrNH, and c) \SrSH. Dashed lines show known rotational decay channels, and identify states that will need to be addressed in order to photon cycle in each species. The asymmetric top states are labeled as $N_{K_a K_c}$. In these molecules, the opposite parity states are not shown due to strict parity selection rules. Note also that these are only the confirmed decay channels; see Appendix~\ref{app:symmetry} for all possible loss channels given the excitations used here. Even when perturbations are taken into account, \SrNH \ requires the fewest rotational repumps per vibrational decay channel.}
    \label{fig:excitations}
\end{figure*}

\subsection{\label{sub:SrOCH3} \texorpdfstring{\SrOCH}{SrOCH3}}
\SrOCH \ is a \ctv \ symmetric top molecule. The ground electronic state is described as a $^2A_1$ representation~\cite{Wormsbecher1982,OBrien1988}. Levels in this state are well-represented by Hund's case (b). Since the end-over-end rotational constant, $B$ (0.084\wn~\cite{Forthomme2011}), is much smaller in SrOCH$_3$ than the symmetry-axis rotational constant, $A$ (5.185\wn~\cite{Forthomme2011}), the rotational levels labeled by $N$ are grouped into ladders of distinct values of $K=|K\s|$, where $K\s$ is the projection of $\textbf{N}$ onto the symmetry axis, $a$; for clarity, we employ a superscript silcrow to distinguish signed angular momentum projections from unsigned ones. The spacing between $K$ manifolds is set by $A-B$. In states with $K \neq 3n$, each $N$ level has a rotational $E$ representation and possesses nearly degenerate opposite-parity states, known as a $K$-doublet, split only by weak hyperfine interactions~\cite{Butcher1993}. Thus any state in these rotationally excited $K$ manifolds will be easily polarized and therefore of interest for EDM or quantum science experiments.

The lowest electronic excited state is described as a $^2E$ representation~\cite{Wormsbecher1982,OBrien1988}, analogous to the $^2\Pi$ state in SrOH. As in SrOH, there is large spin-orbit coupling that splits the manifold into $^2E_{1/2}$ and $^2E_{3/2}$ manifolds; we drive transitions only to the former. This state is Hund's case (a), so that each rotational eigenstate has significant admixtures of distinct $N$ values.

We use the notation $(N,K;J)^{\pm}$ to label rotational states in \SrOCH. In the situation of Hund's case (a) states where $N$ is a bad quantum number, we omit the $N$ label. Likewise, we omit $J$ and parity labels unless necessary.

For molecules in the $C_{3v}$ point group, rotational closure schemes have been previously identified~\cite{Kozyryev2016,Mitra2020,Kozyryev2019}. We reiterate the main points with a slightly modified discussion that emphasizes the structural selection rules in \ctv molecules, which will elucidate the comparison with the $C_{2v}$ and $C_s$ cases. The differentiating factors between rotational closure in linear versus \ctv \ molecules arise for states with non-zero rigid body rotational angular momentum about the $a$-axis, $K_R$, which have no analog in linear species. 

Since there is no electric dipole moment perpendicular to the symmetry axis, rovibronic transitions do not change $K_R$ (up to small perturbations such as the Jahn-Teller and pseudo-Jahn-Teller interactions that couple rotational and other angular momenta \cite{Brazier1989,Marr1996,Paul2019}) and an optical cycle must occur between states with the same value of $K_R$. The parity doublet structure of interest to symmetry violation experiments is not found in the $K=K_R=0$ manifold of \Xctv, so we focus on the optical cycles available with $K_R=1$.

A given \Actv$(N',K')$ component of an eigenstate may decay to \Xctv$(N'',K'\pm1)$ where $N''=N'$ or $N'\pm1$, subject to $N''\geq K''$ and $K''\geq 0$, a double-prime denotes a ground state label, and a single-prime denotes an excited-state label. Ideally, the $N'=K'=0$ state, with $J'=1/2$, would be targeted so that only $N''=K''=1$ can be populated. Clearly, a \Actv$(0,0;1/2)$ state may mix with the \Actv$(1,0;1/2)$ state of the same parity while respecting $J$ and parity conservation. Such a mixing will contribute additional decays to \Xctv$(2,0)$, necessitating a rotational repumping transition as demonstrated in Ref.~\onlinecite{Mitra2020} for the CaOCH$_3$ laser cooling cycle originating from $K''=1$.

We now present a brief heuristic argument for why the $C_{3v}$ molecular symmetry offers no \emph{additional} protection against rotational state mixing in the \Actv($K'=0$) manifold beyond that afforded by angular momentum and parity selection rules. States with distinct $N$ components within the \Actv$(K'=0)$ manifold essentially arise from different combinations of $(N_R',K_R'=1)$ states, where we have denoted the rigid body rotational angular momentum by $N_R$ to distinguish it from the total angular momentum excluding spin, $N$. Since \emph{all} of the $(N_R',K_R'=1)$ rotational states have an $E$ representation in the $C_{3v}$ point group, it follows that there cannot be a symmetry-based prohibition against mixing $N'=J-1/2$ and $N'=J+1/2$ states of the same parity within the $K'=0$ manifold (which has $K_R'=1$). Since the \Actv\ state is Hund's case (a), the symmetry-allowed mixing between $(0,0;1/2)$ and $(1,0;1/2)$ components is of order unity and the $(0;1/2)$ state decays with comparable strength to both \Xctv$(1,1)$ and \Xctv$(2,1)$.

To study the vibrational branching ratios from the excited state used for laser cooling, we drive \Xctv$(1,1;1/2)^{\pm} \rightarrow $\Actv$(0;1/2)^{\mp}$. Here the parity is denoted with $\pm$ ($\mp$) in the ground (excited) manifold because the parity doublets are split by a smaller energy gap than the natural linewidth of the transition. The \Xctv$(1,1)$ state is significantly populated in our CBGB despite having a relatively large absolute rotational energy (owing to the large rotational constant $A=5.2$\wn~\cite{OBrien1988}) because it is the ground rovibronic state of the para-\SrOCH\ isomer~\cite{Mitra2020}.

We also attempted to study SrCH$_3$ as part of this work. However, we were unable to observe the molecule in our CBGB and, as a result, tentatively suggest that SrCH$_3$ be disfavored compared to SrOCH$_3$ for future laser-cooling applications; see App.~\ref{app:srch3} for details.

\subsection{\label{sub:SrNH2} \texorpdfstring{\SrNH}{SrNH2}}
With a slightly lower degree of symmetry, \SrNH belongs to the \cwv \ point group. The ground state is described by the $^2A_1$ representation~\cite{Wormsbecher1983,Bopegedera1987}. As in \SrOCH, this state is Hund's case (b)~\cite{Brazier2000}. Individual rotational levels are again described by $N$. The projection $K$ is no longer a good quantum number, but we can identify states with the additional labels $K_a$ $(K_c)$, denoting the value of $K$ that would describe the state if the molecule were adiabatically deformed to a prolate (oblate) symmetric top. Because $A\gg B\approx C$ in \SrNH , the level structure closely resembles that of a prolate symmetric top with $K_a$ corresponding to $K$~\cite{Brazier2000}. In the relevant case where the symmetry axis is along the $a$ inertial axis, the parity of a state alternates with even vs. odd values of $K_c$ and in the vibronic ground state, the parity is given by $P=(-1)^{K_c}$. The allowed values of $K_c$ are $N-K_a$ and $N-K_a+1$. The states comprising an effective parity doublet are therefore the two allowed states with the same $N$ and $K_a\geq1$ but different $K_c$. These states are inherently split by an amount set by the rotational constant asymmetry $B-C>0$. In \SrNH\, the splitting between opposite-parity states in the $N=K=1$ manifold (in the rigid rotor model neglecting details of hyperfine and spin-rotation structure) is $\sim$130~MHz \cite{Thompsen2000}, large enough to be spectroscopically resolvable but small enough to be easily polarized.

The \Acwv \ state differs qualitatively from the first excited state in SrOH and \SrOCH because in the lower symmetry group there are no doubly-degenerate electronic manifolds. Physically, the degeneracy of the $\Pi$ orbital is broken by the orientation of the hydrogen nuclei, and the in-plane vs. out-of-plane electron orbitals acquire distinct energies. However, there is still a moderately strong spin-orbit interaction, so the \Acwv\ state is described by Hund's case (a) just as in SrOH~\cite{Brazier2000}. Rotational states are labeled by $(N_{K_a, K_c};J)^{\pm}$, with $J$ and parity omitted unless necessary.

Routes to rotational closure for \cwv molecules have been previously identified~\cite{Augenbraun2020}. For $b$-type transitions like \Xt--\At, the optical cycle proceeds on \Xcwv$(1_{11})\leftrightarrow$\Acwv$(0_{00};1/2)^+$. Because parity alternates with $K_c$, a $(1_{10};1/2)^+$ rotational state from any $B_2$ vibronic manifold (possibly but not necessarily \Acwv) has the same $J$ and parity values as the excited laser cooling state. As seen for the case of $C_{3v}$ molecules, the possibility should be considered that such a $(1_{10})$ state could mix with \Acwv$(0_{00})$, leading to rotational leakage channels such as to \Xcwv$(1_{01})$ or \Xcwv$(2_{21})$. However, such a mixing is strictly forbidden in $C_{2v}$ molecules, up to hyperfine or Coriolis-like interactions that couple states of different rotational symmetries (and which are expected to be very weak). Specifically, the $(0_{00})$ rotational state has the $A_1$ representation, while the $(1_{10})$ state has the $B_1$ representation; thus the $(0_{00})$ excited state rotational label must be highly pure. This behavior guarantees much stronger protection against rotational leakage channels than in $C_{3v}$ or (as we will see) $C_s$ molecules. See Appendix~\ref{app:symmetry} for more details. We note that any rotational leakage channel introduced by the weak rotational-symmetry-violating perturbations mentioned above is likely to require repumping for only the strongest vibrational decays, if at all, in order to cycle $\sim$$10^4$ photons.

To study the laser cooling transition, we drive \Xcwv($1_{11};1/2)^{-}\rightarrow$ \Acwv$(0_{00};1/2)^+$. The molecular symmetry enforces a relationship between the rotational and nuclear states, and (as discussed in Appendix~\ref{app:symmetry}) \Xcwv$(1_{11})$ is the ground state of the $I=1$ isomer. Since the nuclear spin is not efficiently changed during buffer gas cooling, this state thus has a significant population in our CBGB despite being at a relatively large energy (set by $A=13.5$\wn~\cite{Thompsen2000}) compared to the absolute ground \Xcwv$(0_{00})$ state.

\subsection{\label{sub:SRSH} SrSH}

With only a single reflection plane, \SrSH \ belongs to the \cs \ point group~\cite{Fernando1991}. Also an asymmetric top, the electronic and rotational structure is similar to \SrNH. However, the point group is smaller, and as a result there are only two allowed representations. The ground electronic state transforms as the $A'$ representation, and is analogous to the ground states of the other molecules studied here~\cite{Fernando1991}. The rotational structure in \Xt\ is the same as in \SrNH, and we employ the same rotational state notation. In the rigid rotor model (i.e., neglecting the full structure including spin-rotation and hyperfine splittings), the nominal parity splitting in the $N=K_a=1$ rotational state manifold proposed for laser cooling of SrSH is $\sim$30~MHz \cite{Halfen2001}.

The lowest electronic excited state of SrSH also transforms as the $A'$ representation, but is again described as a Hund's case (a) state~\cite{Fernando1991,Sheridan2007}. Though the first-order rotational structure of this state is the same as in \SrNH, the reduced symmetry is compatible with rotational state mixing even in $J=1/2$ states. Specifically, in vibronic $A'$ manifolds the rotational states with $K_c=+1$ transform as $A'$ and those with $K_c=-1$ transform as $A''$. As a result, $A'(0_{00};1/2)$ and $A'(1_{10};1/2)$ states have the same rotational symmetry and may, under the influence of perturbations, generically mix with each other. Provided such a mixing exists, the laser cooling scheme suggested in Ref.~\onlinecite{Augenbraun2020} for the predominantly $b$-type \Xcs--\Acs\ transition will be limited by rotational leakage from the \Acs$(1_{10})$ admixture to \Xcs$(1_{01})$ and \Xcs$(2_{21})$ ground states.

The proposed laser cooling transition, \Xcs$(1_{11})^-\rightarrow$\Acs$(0_{00};1/2)^+$, could not be directly excited in our CBGB because of the large rotational energy of \Xcs$(1_{11})$, set by the rotational constant $A=9.71$\wn~\cite{Halfen2001}, and correspondingly low thermal population. In principle, the \Xcs--\Acs\ transition dipole moment contains a small ``$a$-type'' amplitude, so that \Acs$(0_{00})$ can also be populated from \Xcs$(1_{01})$. We therefore drive \Xcs$(1_{01};1/2)^- \rightarrow$\Acs$(0_{00};1/2)^+$ in order to measure rovibrational branching ratios from the proposed laser cooling excited state. We find this transition to be of comparable strength to $b$-type transitions in the \Xcs--\Acs\ band, consistent with the reports of~\cite{Sheridan2007} where strong transitions were observed between $K_a''=0$ and $K_a'=0$ states due to a large excited-state perturbation. Though a concrete physical origin of the perturbation was not identified in that work, and cannot be inferred from our measurements, we interpret these results as supporting the existence of the mixing suggested above between \Acs$(0_{00})$ and \Acs$(1_{10})$. We note that~\cite{Sheridan2007} observed less evidence of perturbation in the excited state of CaSH, highlighting that although mixing between $(0_{00})$ and $(1_{10})$ would be \emph{permitted} in the excited state of any $C_s$ molecule, its actual prevalence must be assessed on a case-by-case basis. We provide additional evidence for this mixing in SrSH, and discuss its implications for laser cooling, in Sec.~\ref{sec:discussion}.

\begin{table}
        \begin{tabular}{ c  c  c }
            \multicolumn{3}{c}{\SrOCH} \\
            \hline
            \hline
            Number & Description & Representation \\
            \hline
            1 & C--H sym. stretch & $a_1$ \\
            2 & CH$_3$ umbrella & $a_1$\\
            3 & C--O stretch & $a_1$ \\
            4 & Sr--O stretch & $a_1$ \\
            5 & C--H asym. stretch & $e$ \\
            6 & Scissor & $e$ \\
            7 & Rock & $e$ \\
            8 & Sr--O--C bend & $e$ \\
            \hline 
            
            \multicolumn{3}{c}{} \\ 
            
            \multicolumn{3}{c}{\SrNH} \\
            \hline
            \hline
            Number & Description & Representation \\
            \hline
            1 & N--H sym. stretch & $a_1$ \\
            2 & NH$_2$ bend & $a_1$\\
            3 & Sr--N stretch & $a_1$ \\
            4 & Sr--N--H out of plane bend & $b_1$ \\
            5 & N--H asym. stretch & $b_2$ \\
            6 & Sr--N--H in plane bend & $b_2$ \\
            \hline 
            
            \multicolumn{3}{c}{} \\ 
            
            \multicolumn{3}{c}{\SrSH} \\
            \hline
            \hline
            Number & Description & Representation \\
            \hline
            1 & S--H stretch & $a'$ \\
            2 & Sr--S stretch & $a'$\\
            3 & Sr--S--H bend & $a'$ \\
            \hline
        \end{tabular}
        \caption{Ground state vibrational mode labels and representations for each of the species studied here.}
        \label{tab:statesym}
\end{table}

\section{\label{sec:analysis} Results}

\begin{figure}
    \centering
    \includegraphics[width = \columnwidth]{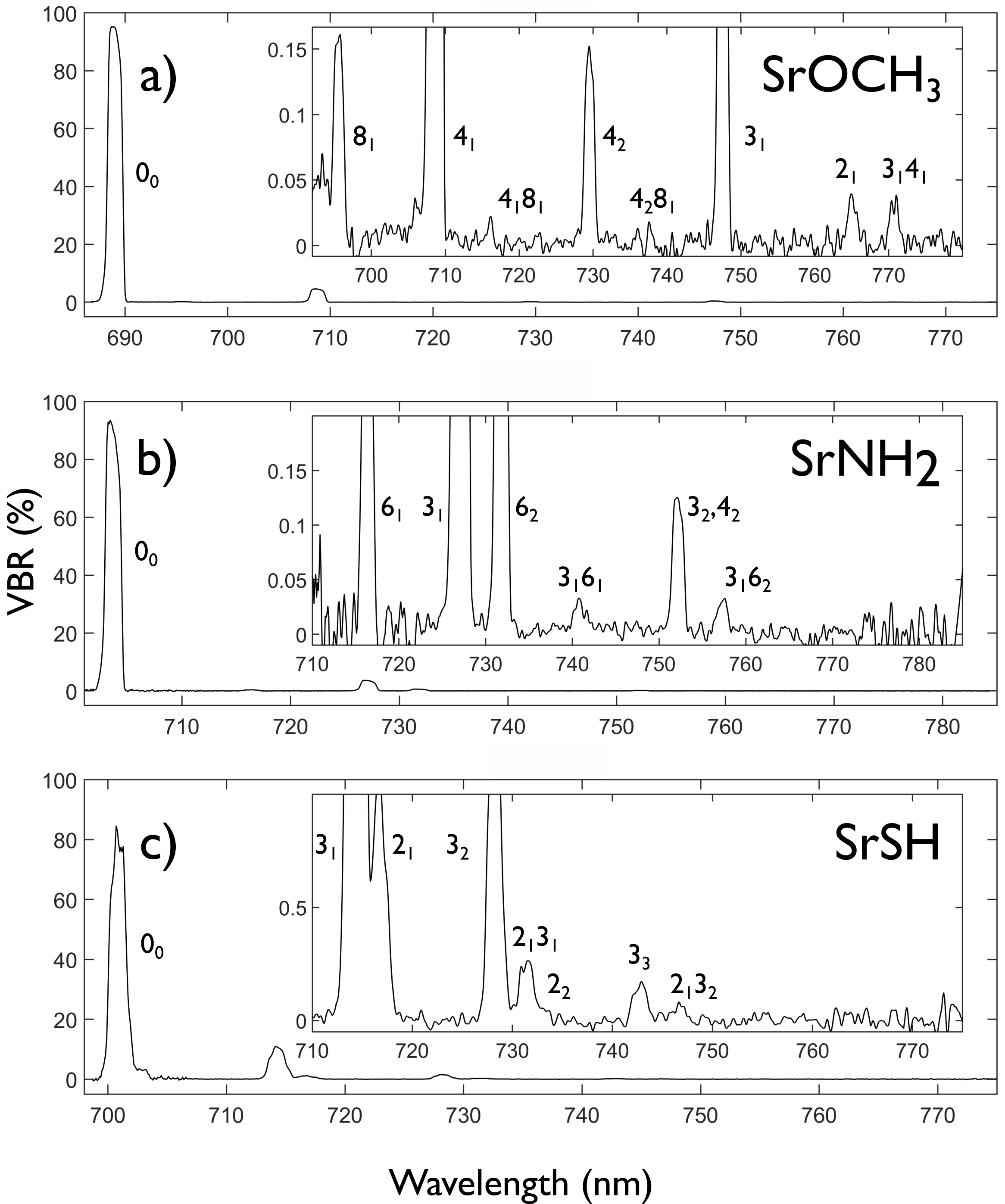}
    \caption{VBR data for (a) \SrOCH, (b) \SrNH, and (c) \SrSH. Insets show the same spectra at higher resolution. Accompanying VBRs are found in Tables~\ref{tab:SrOCH3-VBR-peak}--\ref{tab:SrSH-VBR-peak}. States are labeled according to the modes in Table \ref{tab:statesym}. State labels separated by commas indicate transitions to different states that are unresolved by the spectrometer.}
    \label{fig:vbrs_all}
\end{figure}

By driving transitions to the relevant rovibrational levels in the first excited electronic states of \SrOCH, \SrNH, and \SrSH, we record VBRs out of these states at or below the $\sim$0.1\% level. The resulting dispersed fluorescence spectra can be seen in Fig.~\ref{fig:vbrs_all}. 

We use the calculated vibrational energies and branching ratios to help identify each decay channel. We denote the $i$-th vibrational mode $v_i$, identified for each molecule of interest in Table~\ref{tab:statesym}. Vibrational energy levels are labeled according to the convention $i_{n(i)}j_{n(j)}\cdots$ to denote the state with $n(i)$ quanta of excitation in mode $v_i$, and so on. Modes with $n(i)=0$ are omitted, and the special case of the vibrational ground state is denoted $0_0$. We do not measure or assign rotational branching ratios to vibronic levels where multiple \textit{rotational} levels are populated. Peak intensities are computed by integrating the signal over the width of a feature. The resulting calculated VBRs can be found in Tables~\ref{tab:SrOCH3-VBR-peak}--\ref{tab:SrSH-VBR-peak}.

\section{\label{sec:discussion} Discussion}

\subsection{\label{subsec:dsroch3} \texorpdfstring{\SrOCH}{SrOCH3}}

We measure the vibrational branching ratios of \SrOCH \ over a range of 80~nm. We reach an ultimate VBR sensitivity over this range of $\sim$0.01\%. We find qualitatively worse scaling of vibrational branching than in the isoelectronic linear molecule SrOH; see Table~\ref{tab:SrOCH3-VBR-peak}. We note two contributions to this difference.

First, the decay to a single quantum of the Sr--O--C bending mode ($8_1$) is about an order of magnitude larger than to the analogous $(010)$ state of SrOH. This bending state is also accompanied by higher-probability decays to some combination states ($4_1 8_1$ and $ 4_2 8_1$). Two vibronic coupling mechanisms contribute to these decays. One is the Jahn-Teller effect (JTE), which mixes $0_0$ and $8_1$ in the \At \ state, making these decays no longer vibronically forbidden~\cite{Bunker2006}. The other mechanism is vibronic coupling between the \At$^2E$ states and the $\tilde{B}^2 A_1$ state. This is essentially the same as the ``direct vibronic coupling'' (DVC) mechanism \cite{Zhang2021,Zhang23} in CaOH and SrOH that borrows the intensity for the VBRs to (010) states. In the following we also refer to this vibronic coupling mechanism as the DVC mechanism. 
 
These effects similarly contribute to enhancing the strength of decays in \CaOCH \ to the $8_1$ mode compared to $(010)$ in CaOH~\cite{Baum2021,Paul2019, Augenbraun2021a}. In the present computational treatment, the inclusion of spin-orbit coupling has been shown to quench the JTE; the first excited state with spin-orbit coupling has the $C_{3v}$ structure as a minimum on the potential energy surface. We have included the DVC contribution via perturbation theory using the formulation reported in Ref. ~\onlinecite{Zhang23}. The VBR to $8_1$ thus computed agrees reasonably well with the measured value. The VBR to $8_1$ in SrOCH$_3$ is larger than that to the first excited bending mode in SrOH. This is readily attributed to that the value of around 120 cm$^{-1}$ for the linear vibronic coupling constant \cite{Ichino09} between the \At$^2 E$ and $\tilde{B}^2 A_1$ states in SrOCH$_3$ is larger than the value of 70 cm$^{-1}$ between the \At$^2\Pi$ and $\tilde{B}^2\Sigma$ states in SrOH \cite{Lasner2022}.

The second cause of increased branching channels is simply that there are more modes (of both allowed and nominally forbidden symmetry) to decay to than in smaller molecules. In particular, in addition to the $a_1$ symmetry Sr--O stretch ($v_4$), \SrOCH \ also has an $a_1$ symmetry C--O stretch ($v_3$). We find decay to both the $3_1$ and $3_1 4_1$ states above our measurement sensitivity. Two other $a_1$ vibrational modes also exist, the $\textrm{CH}_3$ ``umbrella'' mode ($v_2$) and the C--H symmetric stretch ($v_1$). Near the noise floor we tentatively assign a small decay to $2_1$, though as expected it is suppressed compared to the modes more closely coupled to the Sr atom. The decay to $1_1$ lies beyond our measured wavelength range, but calculations predict the decay to this mode to be below our measurement sensitivity of $\sim$0.01\%. Decays to the three remaining modes in the molecule---namely, the C--H asymmetric stretch ($v_5$), $\textrm{CH}_3$ scissor ($v_6$), and $\textrm{CH}_3$ rock ($v_7$)---would be accompanied by infrared fluorescence outside of our measurement range but are all nominally symmetry forbidden. Our computational work supports the negligibility of these modes at the sensitivity of the present work. Thus several (though not all) modes in \SrOCH\ without any analog in smaller species such as SrOH or SrSH contribute relevant vibrational decays around the $0.01\%$ level.

\begin{table}
    \begin{center}
    \begin{tabular}{|c||c|c||c|c|}
        \multicolumn{5}{c}{\SrOCH \ \Actv \ $0_0$}\\
        \hline
         & \multicolumn{2}{c||}{Energy (cm$^{-1}$)} & \multicolumn{2}{c|}{VBR (\%)} \\
        \hline
        State & Calc. & Exp. & Calc. & Exp. \\
        \hline
        $0_0$     & 0    & 0         & 92.688    & 94.7(2)    \\
        $8_1$     & 144  & 138(5)    & 0.145     & 0.18(8)    \\
        $8_2$     & 286  & ---       & 0.066    & $<$0.02 \\
        $4_1$     & 406  & 404(2)    & 6.369    & 4.5(2)     \\
        $4_1 8_1$ & 553  & 547(3)    & 0.009     & 0.017(7)   \\
        $4_2$     & 810  & 807(2)    & 0.123    & 0.145(10)  \\
        $4_2 8_1$ & 960  & 965(5)    & 0.000     & 0.010(6)   \\
        $3_1$     & 1177 & 1138(2)   & 0.509    & 0.38(2)    \\
        $2_1$     & 1472 & 1446(4)   & 0.055    & 0.037(7)   \\
        $3_1 4_1$ & 1582 & 1542(2)   & 0.023    & 0.033(6)   \\
        $2_1 4_1$ & 1878 & ---       & 0.005    & ---        \\
        $3_2$     & 2339 & ---       & 0.002    & ---        \\
        $1_1$     & 2799 & ---       & 0.006    & ---        \\
        \hline
    \end{tabular}
    \end{center}
    \caption{Predicted and observed vibrational state energies and branching ratios from the \SrOCH \ \Actv\ ground vibrational state. The predicted decays to $2_1 4_1$, $3_2$, and $1_1$ lie outside of the measured wavelength range toward the infrared.}
    \label{tab:SrOCH3-VBR-peak}    
\end{table}

One surprise in the spectrum is the lack of an observed $8_2$ decay. Calculations predict that it should occur around the $0.07\%$ level, as it is vibronically allowed. However, we see no such decay, and bound the decay to $<0.02\%$. We know of no technical reason that a peak larger than this would not appear in our dataset.

In all, we find at least 9 vibrational levels populated above $0.01\%$ probability, compared to 5 in SrOH. The measurements and predictions are in generally good agreement. We therefore tentatively extrapolate from the computed results that approximately 11 vibrational states will be populated in a photon cycle with $\gtrsim$15,000 scatters, compared to 8 in SrOH~\cite{Lasner2022}. In addition, as noted in Sec. \ref{sec:molecules}, most of these vibrational repumpers will require addressing two well-separated rotational states, increasing the experimental challenge significantly further. Since the production rates of SrOH and \SrOCH\ are comparable, these factors unambiguously indicate a much more challenging laser cooling scheme for \SrOCH. Nevertheless, as the rotational closure is well understood and the VBR measurements identify the relevant peaks to the $\sim$0.01\% level, we expect \SrOCH\ to be realistically fully laser coolable with sufficient effort.

\subsection{\label{subsec:dsrnh} \texorpdfstring{\SrNH}{SrNH2}}

\begin{table}
    \begin{center}
    \begin{tabular}{|c||c|c||c|c|}
        \multicolumn{5}{c}{\SrNH \ \Acwv \ $0_0$}\\
        \hline
         & \multicolumn{2}{c||}{Energy (cm$^{-1}$)} & \multicolumn{2}{c|}{VBR (\%)} \\
        \hline
        State & Calc. & Exp. & Calc. & Exp. \\
        \hline
        $0_0$     & 0    & 0         & 94.024  & 95.1(2) \\
        $6_1$     & 260  & 252(4)    & 0.351     & 0.35(3) \\
        $3_1$     & 454  & 459(1)    & 4.627    & 3.7(2)  \\
        $6_2$     & 531  & 547(2)    & 0.869    & 0.63(3) \\
        $3_1 6_1$ & 699  & 717(9)    & 0.019     & 0.04(1) \\
        $3_2$     & 904  & 918(4)    & 0.048    & 0.10(1) \\
        $4_2$     & 888  & $\sim$918 & 0.012    & 0.03(1) \\
        $3_1 6_2$ & 965  & 1007(2)   & 0.030    & 0.03(1) \\
        $6_4$     & 1083  & ---      & 0.002    & $<$0.03 \\
        $2_1$     & 1540 & ---       & 0.014    & ---     \\
        $1_1$     & 3331 & ---       & 0.003    & ---     \\        
        \hline
    \end{tabular}
    \end{center}
    \caption{Predicted and observed vibrational state energies and branching ratios from the \SrNH \ \Acwv\ ground vibrational state. We tentatively assign the observed 918\wn\ peak to the sum of unresolved decays to $3_2$ and $4_2$. The ratio of VBRs to these two components is assumed to match the calculated ratio. The expected decays to $2_1$ and $1_1$ lie outside of the measured wavelength range.}
    \label{tab:SrNH2-VBR-peak}    
\end{table}

We measure VBRs for the \SrNH \ \Acwv \ $0_0$ state over a similarly wide range of $\sim$80~nm. Due to lower production, however, our ultimate sensitivity is $\sim$0.05\%. We find a similar number of states populated at our ultimate sensitivity as in SrOH (see Table~\ref{tab:SrNH2-VBR-peak}). However, there are several important differences between \SrNH\ and its linear analog. Most notably, the Sr--NH$_2$ bending motions, which in a linear species are doubly degenerate, are separated by $>$100\wn\ and belong to different 1D representations. Decay to the in-plane vibrational bending excitation, $6_1$, can be induced by vibronic perturbations between $\A B_2 0_0$ and $\C A_1 6_1$, owing to the $b_2$ symmetry of the $v_6$ mode. This decay is much stronger than the analogous $(010)$ decay in SrOH. The out-of-plane $v_4$ mode is predicted to be populated in the $4_2$ level near our measurement resolution, though we do not observe a corresponding peak. Thus, despite the lack of bending degeneracy, the lower symmetry of \SrNH\ compared to SrOH does not seem to induce more loss channels at the $0.05 \%$ level. 

Another difference from SrOH is that the $\textrm{NH}_2$ bend is predicted to be populated around the $0.01\%$ level, though it is outside of our measured wavelength range. This is a significantly higher level than ligand modes in SrOH, and indicates a less polar bond, which is supported by spectroscopic analysis \cite{Bopegedera1987}. The branching ratio to the N--H stretch mode is predicted to be around $0.003\%$, smaller than that of the C--H stretch in CaOCH$_3$. This is likely not limiting for a target photon budget of $\sim$$10^4$ scatters. Overall, the loss channels introduced by the amine group are manageable though nonzero.

In addition to these structural differences, there are also several resonances that appear to perturb the ground state vibrational manifold substantially. In particular, the $3_1$ and $6_2$ states are close in energy and of the same symmetry, and so are mixed by a strong Fermi resonance. Overtones and combination bands will also be mixed (e.g., $3_2$ and $3_1 6_2$). Consider the transitions to the $3_1$ and $6_2$ states as examples. The computed branching ratio to the $6_2$ state within the harmonic approximation is less than 0.1$\%$. We thus infer that the branching ratio of 0.6\% for the transition to $6_2$ observed here predominantly arises from the mixing with the $3_1$ state due to the Fermi resonance. The VPT2 calculation that explicitly diagonalizes the $2\times 2$ matrix expanded by the $3_1$ and $6_2$ harmonic oscillator wave functions shows that the $6_2$ state has around $15\%$ contribution from the $3_1$ harmonic oscillator wave function. This gives rise to the $\sim$1:5 intensity ratio between the decays to the $6_2$ and $3_1$ states. Since these states would likely be populated in $10^4$ optical cycles even in the absence of Fermi resonances, these ground state resonances should not substantially affect the complexity of laser cooling the molecule. A similar Fermi resonance appears in CaOH between the $(100)$ and $(020)$ modes and does not significantly impact the number of states needed for full photon cycling~\cite{Baum2021,Vilas2022}.

We also note that the present calculations may not have captured the resonances among the overtones accurately. The computed levels for $3_1$ and $6_1$ agree well with the measured values. However, the computations substantially underestimate the level positions for the $3_1 6_1$ and $3_1 6_2$ states. Future computational work will be focused on improving the potential energy surfaces and performing variational calculations of vibrational structures to go beyond VPT2.

We find only 7 vibrational states populated at the $\sim$0.05\% level, slightly more than in SrOH. Extrapolating to the $\gtrsim$15,000 photon scatter level using computed branching ratios, only 9 vibrational states are expected to be populated. At the same branching probability level there are 8 populated vibrational levels in SrOH. However, because every vibrational decay is rotationally closed in \SrNH (unlike for the bending mode excitations in SrOH), remarkably we tentatively anticipate one fewer repumping laser to be necessary to achieve the same degree of rovibronic closure. Thus \SrNH \ appears a strong choice for full laser cooling in a future precision measurement, as it is significantly simpler both rotationally and vibrationally than \SrOCH, while possessing long-lived parity doublet structure in the ground state. 

\subsection{\label{subsec:dsrsh} \SrSH}

\begin{table}
    \begin{center}
    \begin{tabular}{|c||c|c||c|c|}
        \multicolumn{5}{c}{\SrSH \ \Acs \ $0_0$}\\
        \hline
         & \multicolumn{2}{c||}{Energy (cm$^{-1}$)} & \multicolumn{2}{c|}{VBR (\%)} \\
        \hline
        State & Calc. & Exp. & Calc. & Exp. \\
        \hline
        $0_0$     & 0   & 0         & 88.383    & 85.2(5) \\
        $3_1$     & 268 & 268(2)    & 10.206    & 10.9(5)\\
        $2_1$     & 323 & 315(4)    & 0.629    & 1.85(9) \\
        $3_2$     & 540 & 536(2)    & 0.499    & 1.51(8) \\
        $2_1 3_1$ & 604 & 596(3)    & 0.113    & 0.29(2) \\
        $2_2$     & 634 & $\sim$623 & 0.102    & 0.07(4) \\
        $3_3$     & 815 & 805(2)    & 0.013 & 0.17(3) \\
        $2_1 3_2$ & 889 & 875(2)    & 0.008 & 0.07(3) \\
        $2_2 3_1$ & 928 & ---    & 0.010 & $<0.03$ \\
        $1_1$     & 2895 & ---    & 0.036 & --- \\
        \hline
    \end{tabular}
    \end{center}
    \caption{Predicted and observed vibrational state energies and branching ratios from the SrSH \Acs\ ground vibrational state. A shelf toward the red side of the peak arising from decays to $2_1 3_1$ is tentatively assigned to $2_2$; however, a reliable energy estimation for $2_2$ cannot be made due to the partially unresolved spectrum. The predicted decay to $1_1$ lies outside of the measured wavelength range.}
    \label{tab:SrSH-VBR-peak}    
\end{table}

In SrSH we see evidence of decay to \textit{all} $K_a$ manifolds allowed by selection rules on $J$ from the \Acs$(0_{00};1/2)^-$ excited state. We find this well-explained by the symmetry-allowed mixing suggested previously between \Acs$(0_{00})$ and \Acs$(1_{10})$. Sheridan \textit{et al.} also suggested a large perturbation (of unclear fundamental origin) in SrSH that mixes $K_a'=0$ and $K_a'=1$ states. 

To test the effect of the perturbation more rigorously, we recorded the fluorescence spectrum for excitations to several rotational levels in the \Acwv $0_0$ state. We narrowed the spectrometer slit to $\sim$$5~\mu$m to resolve the decays to distinct $K_a$ manifolds, which are spaced on a scale set by the a-axis rotational constant, $A=9.71$\wn, and thus resolvable~\cite{Halfen2001}. Branching to different $N$ and $J$ levels within each $K_a$ manifold remains unresolved.

\begin{figure}
    \centering
    \includegraphics[width = \linewidth]{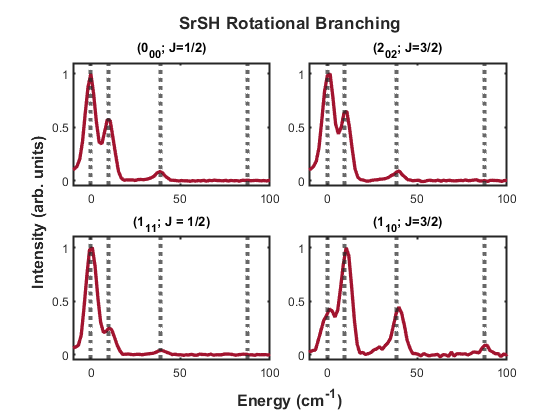}
    \caption{Comparison of rotational branching for four rotational levels in the \SrSH \ \Acs \ $0_0$ state to the \Xcs \ $0_0$ state. Each peak corresponds to a decay to a different $K_a$ level in the ground state. Dashed lines indicate the energies of $K_a$ levels using the measured $A$ constant \cite{Brazier2000}. Note that four decays only happen for $K_a' = 1$ \textit{and} $J' = 3/2$. The other states show only three decays.}
    \label{fig:rotationalbranching}
\end{figure}

In particular, we measured the decay patterns to the \Xcwv $0_0$ vibrational state for excitations to $J' = 1/2$ and $J' = 3/2$ states in both $K_a' = 0,1$ manifolds. The results can be seen in Fig.~\ref{fig:rotationalbranching}. There are clearly more decay channels from the nominal \At$(1_{10};3/2)$ state than from either of the \At$(0_{00};1/2)$ or \At$(1_{11};1/2)$ states, indicating a dependence of $K_a$ branching ratios on $J$. At the simplest level, this behavior occurs because an excited $J'=1/2$ state can mix with other $J'=1/2$ states, implying at most $N'=0,1$ components. Since $K_a\leq N$, consequently a $J'=1/2$ state has at most contributions from the $K_a'=0,1$ manifolds. For a $b$-type transition, these components can populate $K_a''=0,1,2$ upon decay. On the other hand, a $J'=3/2$ state can have components with $N'=1,2$ and therefore $K_a'=0,1,2$, which are allowed to decay via $b$-type transitions to $K_a''=0,1,2,3$.

Though parity and angular momentum selection rules would allow \At$(2_{02};3/2)$ to mix with \At$(2_{20};3/2)$, the decay spectrum of \At$(2_{02};3/2)$ does not show a significant decay to $K_a''=3$, which would be expected from any \At$(2_{20};3/2)$ admixture. Thus not \emph{every} pair of states with the same $J$ and parity have order-unity mixing even in the presence of this perturbation. 

Clearly, the admixture of the ``leaky" rotational state \Acs$(1_{10};1/2)$ into the nominal optical cycling excited state, \Acs$(0_{00};1/2)$, is of order unity and will therefore meaningfully affect decays to all higher vibrational branches in \Xcs\ as well. This behavior can be contrasted with any rotational leakage channels that could plausibly be induced in \SrNH\ by Coriolis or hyperfine interactions, where few if any vibrational states should require rotational repumping. The three $K_a$ manifolds populated in each of the SrSH vibrational decays are also too widely spaced ($\gtrsim$300~GHz) from the optical cycle origin to be bridged by typical frequency modulation techniques.

Since the perturbations indicate that laser cooling this species is likely exceedingly difficult, we only measure VBRs to $\sim$0.1\% and examine a fluorescence wavelength range of 60~nm. 
Nevertheless, we can still analyze the VBRs, compare the computations and experiment, and compare to the other species studied here. We find poor vibrational branching for photon cycling, with 6--7 states populated above $0.1\%$---significantly worse than SrOH, \SrOCH, or \SrNH. This comparison is well-explained by the difference in structure of the vibrational modes compared to the other species. Particularly of note, the \cs \ point group has only two representations, both one-dimensional. 
All vibrational modes are $a'$ symmetry. Furthermore, the most relevant modes---the Sr--S--H bend, $v_2$, and the Sr--S stretch, $v_3$---are close in energy and thus decays to one are typically accompanied by decays to the other due to mixing between vibrational states. This effectively introduces a similarly strong bend or combination mode for every Sr--S stretch, which matches the factor of $\sim$2 difference in number of populated decays at the $0.1\%$ level compared to SrOH.

As shown in Table \ref{tab:SrSH-VBR-peak}, the computed vibrational frequencies agree well with the measured ones. The double harmonic approximation tends to underestimate the VBRs to the overtones of the Ca--S stretching mode. This is consistent with the observation for the Ca--O stretching modes of CaOH \cite{Zhang23}. In addition, we mention that the VBR to the S--H stretch mode, $1_1$, in SrSH is predicted to be 0.036\%, an order of magnitude larger than that for the C--H stretch in SrOCH$_3$ and the N--H stretch in SrNH$_2$. This is also unfavorable for laser cooling.

We thus attribute both the poor rotational and vibrational branching in \SrSH \ to the low symmetry of the species. In particular, the absence of higher symmetry is what allows the $K_a$ mixing to occur in the \Acs \ state, as such a mixing is forbidden in higher-symmetry analogs (specifically, in $C_{2v}$ species). Similarly, mixing between all vibrational states is symmetry-allowed, unlike in higher-symmetry species, substantially increasing the number of significant decays beyond SrOH. Given that a single vibrational repumper requires at least three widely-spaced laser frequencies to address all populated $K_a''$ manifolds, and many more vibrational states would be populated in an optical cycle of sufficient depth to fully laser cool the molecule, these findings 
indicate that molecules in the lowest symmetry groups ($C_s$ and $C_1$) pose significant challenges and are less favorable candidates
for laser cooling experiments unless there is a strong specific motivation to use such a species (for example, many chiral molecules, including SrOCHDT~\cite{Augenbraun2020}, belong to the $C_1$ point group). However, such a motivation is not apparent for precision measurement applications such as EDM experiments, where these lowest symmetry groups possess states of only similar polarizability and lifetime as $C_{2v}$ and $C_{3v}$ molecules, which are likely simpler to control.

\section{\label{sec:conclusion} Conclusion}

We measure the vibrational branching ratios of three nonlinear strontium-containing molecules: \SrOCH, \SrNH, and \SrSH. Each of these species has sensitivity to BSM physics searches similar to SrF and SrOH. The lower structural symmetry of these species offers parity doublets in the vibronic ground state, thus combining the polarizability in small electric fields and parity doublets of linear triatomics with the long lifetime of rotational states in diatomics. We measure VBRs and further elucidate to what degree these nonlinear molecules are laser coolable.

There is evidence in the literature of severe perturbation in higher electronic states for all three molecules~\cite{Forthomme2011,Brazier2000,Sheridan2007}, suggesting that the simplest path to photon cycling is likely through the \At \ states. Side-by-side comparisons of laser coolability as a function of molecular point group are studied for these excited states in particular.

We identify complications with rotational closure related to the respective symmetry groups. The low symmetry of \SrSH \ allows perturbations that result in $\geq 3$ rotational decays of comparable strength for each vibrational excitation. Symmetry protections prevent these perturbations from appearing at low order in \SrOCH \ and \SrNH. Conversely, there is one fewer symmetry-allowed rotational loss channel for the laser cooling excitation in \SrNH \ than in \SrOCH due to the lower symmetry of the former. 

To asses vibrational loss channels, we measure the VBRs out of each molecule's \At \ excited state to at least the $0.1\%$ level. We find that the number of vibrational loss channels at this precision are similar between SrOH, \SrOCH, \ and \SrNH. \SrSH \ is found to have significantly more loss channels than SrOH. All loss channels for the three molecules are measured to at least the $0.1\%$ level, sufficient to implement one-dimensional laser cooling in the style of Refs.~\onlinecite{Mitra2020,Augenbraun2020Sisyphus}. Full 3D laser cooling in the typical configuration~\cite{Vilas2022} will require $\sim$$10\times$ higher-resolution VBR measurements for \SrOCH and \SrNH, and $\sim$$100\times$ higher-resolution for \SrSH. 

Looking toward a future polyatomic molecule for the most precise BSM searches, the choice of species to pursue further is clear. In particular, the simplicity of rotational closure in the \cwv \ point group and controlled vibrational branching at the $0.05\%$ level make \SrNH \ the most promising candidate for a next-generation search for the electron electric dipole moment. Further investigation of the \SrNH \ VBRs to the $10^{-5}$ level is thus warranted to identify whether there are any perturbations that appear at higher resolution, which might prevent straightforward full laser cooling and control of the species. These measurements will be possible with the addition of Sr chemical enhancement light \cite{OBrien1988,Brazier2000,Jadbabaie2020} and spin-rotational closure. Such work would enable full laser cooling of \SrNH and make possible an EDM experiment in a long-lifetime conservative trap, reducing the volume over which a high degree of systematic control is necessary.

We expect very similar behavior for other metal containing molecules (e.g. \RaNH, \YbNH, or \BaNH) that have significantly higher BSM sensitivity~\cite{Zhang23}, though spin-orbit-related complications will likely affect the electronic structure of these heavier species. Investigations into the effect of increasingly heavy optical cycling centers on laser cooling in such nonlinear species is therefore highly motivated.
\\
\begin{acknowledgments}
We are grateful to Benjamin Augenbraun for many ideas and help initiating the project. We thank Arian Jadbabie and Nathaniel Vilas for many discussions about molecular structure. Experimental work was done at the Center for Ultracold Atoms (an NSF Physics Frontiers Center) and supported by Q-SEnSE Quantum Systems through Entangled Science and Engineering (NSF QLCI Award No. OMA-2016244) and AFOSR (Award No. FA9550-22-1-0288). Computational work at Johns Hopkins University was supported by the National Science Foundation, under Grant No. PHY-2309253. 
\end{acknowledgments}

\appendix

\section{Attempts to produce \texorpdfstring{SrCH$_3$}{SrCH3}}\label{app:srch3}

As mentioned in Sec.~\ref{sub:SrOCH3}, we were unable to observe SrCH$_3$ in our CBGB using either chloromethane or methane as a reactant gas. This molecule has been observed in room temperature sources~\cite{Dick2006}, though even in that work it was unobserved without Sr atom excitation light present. It has never been observed in cryogenic sources. Further work is needed to identify whether the production is merely suppressed by lack of atomic excitation or whether at cryogenic temperatures production is suppressed via some other mechanism (which would preclude the use of the molecule at all in a laser cooling experiment). Even in the former case, all other laser cooled species that report enhanced production from atomic excitation have some non-negligible production without it, and it thus seems questionable whether production could match the output of SrOH or the other species studied here. In light of this, we suggest that this molecule does not appear especially promising for EDM measurements due to a substantial challenge with production.

\section{Symmetry analysis for \texorpdfstring{$C_s$}{Cs} and \texorpdfstring{$C_{2v}$}{C2v} rovibronic branching}\label{app:symmetry}

In this Appendix, we expand upon the ideas presented in Sec.~\ref{sec:molecules} regarding the consequences of molecular symmetry point groups on the rotational closure in $C_s$ and $C_{2v}$ molecules. For each of these point groups, we present an overview of the relevant representation theory with connections to the special cases of SrNH$_2$ and SrSH. We also briefly treat the situation for $C_1$ molecules (for example, SrOCHDT) for comparison. We do not treat $C_{3v}$ molecules here, as the essential concepts specific to that point group are already addressed in the literature~\cite{Mitra2020}. We first discuss generic considerations of molecular structure, and then treat each point group individually. All representation theory results are derived from basic principles outlined in~\cite{Bunker2006} but are developed here in detail for clarity and reference.

\subsection{Transition selection rules and perturbations}\label{sec:transition-rules}

\begin{table}
    \centering
    \begin{tabular}{c|ccc}
                 &  $\Delta K_a$ & $\Delta K_c$ & Exceptions\\
              \hline
        $a$-type & 0             & $\pm$1       & $\Delta N\neq 0$ for $K_a'\rightarrow K_a''=0$\\
        $b$-type & $\pm$1        & $\pm$1       & \\
        $c$-type & $\pm$1        & 0            & $\Delta N\neq 0$ for $K_c'\rightarrow K_c''=0$\\
    \end{tabular}
    \caption{Rotational transition rules for asymmetric tops, up to weak transitions allowed by perturbations (e.g., those that make $N$, $K_a$, or $K_c$ imperfect state labels). In a structureless rigid rotor, $|\Delta N|\leq 1$.}
    \label{tab:rotational-rules}
\end{table}

For all asymmetric top molecules, the nominal transition selection rules are shown in Table~\ref{tab:rotational-rules}; see Ref.~\onlinecite{Augenbraun2020}. The $a$-type, $b$-type, and $c$-type selection rules apply for cases in which the transition dipole moment is along the $a$, $b$, and $c$ axis, respectively. We denote a transition dipole moment component along the $i$ axis (where $i=a,b,c$) by $T_i$. In realistic cases, a transition dipole moment may have nonzero components along multiple principal inertial axes. For example, the \Xcs--\Acs\ transition dipole moment in SrSH is predominantly along the $b$ axis but also has a small component along the $a$ axis. We also note that a given energy eigenstate might have contributions from multiple $(N_{K_a K_c})$ components. We warn the reader that in the literature, a single $(N_{K_a K_c})$ label may be employed to label an energy eigenstate based on the dominant rotational basis component, even if a small admixture of other rotational basis components is present. Up to hyperfine structure and the influence of external fields, $J$ is a perfectly good quantum number, and for doublet molecules $N=J\pm1/2$, constraining the identity of $(N_{K_a K_c})$ basis states that can mix.  In $C_s$, $C_{2v}$, and $C_{3v}$ molecules the parity of a state is also pure even in the presence of realistic perturbations. All dipole moment and rotational basis components will contribute to the transition strengths between two energy eigenstates according to Table~\ref{tab:rotational-rules}.

We now provide a general overview of vibronic decays. In the Born-Oppenheimer approximation (BOA), the vibronic transition dipole operator acts only on the electronic state, and the transition probability from an initial vibrational state $|i\rangle$ to a final vibrational state $|j\rangle$ is proportional to $|\langle j|i\rangle|^2$ (i.e., governed by Franck-Condon factors). This inner product is non-zero only for a final vibrational state with the same representation as the initial vibrational state. Thus in the BOA, the representation of the vibrational state is unchanged during decay.

However, the vibronic representation is obtained from the product of electronic and vibrational representations. In certain cases, interactions can mix states of different electronic and vibrational symmetries but the same vibronic symmetry. This kind of vibronic coupling is a signature BOA breakdown. Typically, BOA-forbidden decays are weak (e.g., $\sim$$10^{-3}$ branching fraction) but not necessarily negligible at the level of vibrational closure required for deep laser cooling. For this reason, here we will rely on the vibronic representation (rather than electronic and vibrational representations separately) to determine better-respected selection rules on transitions.

In the same manner that states of different electronic and vibrational representations (but the same vibronic representation) can mix via vibronic coupling, it is possible for states of different vibronic representations to mix. The first notable way in which mixing between states with different vibronic representations occurs is Coriolis interactions, which couple rigid body rotation with some other angular momentum (such as a vibrational or electronic angular momentum). These interactions can mix states only of the same total \emph{rovibronic} representation. Such perturbations will typically be relatively small, and their effects (if any) on optical cycling have already been discussed in Ref.~\onlinecite{Augenbraun2020}. The second notable way in which states with different vibronic representations mix is spin-orbit interactions, via a term in the Hamiltonian like $A^{\rm{SO}}L_i S_i$. This operator has the same representation as the angular momentum $J_i$, which is not generically the totally symmetric representation and therefore can produce mixed vibronic representations for an energy eigenstate. Such a term is possible because the full symmetry group of a spin-doublet molecule is an electron spin double group~\cite{Bunker2006} rather than the molecular symmetry group that is more conventionally used in the literature (including here). We note that for molecules like SrSH and SrNH$_2$, because the ligand quenches orbital angular momentum (in low-$J$ states), we can expect spin-orbit interactions to have a smaller influence on electronic state mixing than in species like SrOH or SrOCH$_3$ with $\sim$$\hbar$ of electronic orbital angular momentum around the molecular symmetry axis. Insofar as such effects do mix states of different overall vibronic symmetry, they play a role in laser cooling analogous to the Coriolis effects mentioned above.

All vibronic-symmetry-violating perturbations are expected to be weak and to affect rotational branching only at a low level, if at all (e.g., after scattering $\sim$$10^3$--$10^4$ photons or more). As a result, any rotational leakage channels introduced by such effects will likely be important only for the vibrational states populated most frequently in an optical cycle, and can be addressed if needed by one or a few rotational repumps even in a deep optical cycle. Another perturbation that could conceivably be important in a sufficiently deep optical cycle is hyperfine interactions, which could mix states of different nuclear and rotational representations, but the same total molecular state representation. Because hyperfine interactions are small ($\lesssim1$~MHz) in molecules like SrSH and SrNH$_2$, where the nuclear-spin-bearing hydrogen atoms are far from the metal-centered valence electron, here we treat them as having a negligible effect on rotational state purity. 

To summarize, in the following treatment of symmetry-allowed and symmetry-forbidden transitions in $C_s$ and $C_{2v}$ molecules, we assume that there may be interactions that mix states of different electronic and vibrational symmetries, provided the \emph{vibronic} symmetry remains pure. Specifically, we neglect any possible effect of Coriolis or spin-orbit interactions that could, in principle, mix states of different vibronic symmetry. We also neglect any possible effect of hyperfine interactions, which could mix states of different rotational symmetry. Otherwise, we allow for the possibility of mixing rotational state components $(N_{K_a K_c})$ in the excited manifold. We will not consider the effect of rotational state mixing in the ground electronic manifold. Such mixings have been considered in Ref.~\onlinecite{Augenbraun2020}, and can result in rotational leakage on the order of $\sim$$10^{-5}$. This is likely negligible for realistic experiments, and could easily be addressed by microwave or optical repumping schemes for any frequently-populated vibrational states in the ground electronic manifold. Vibrational states populated later in the optical cycle are even less likely to require repumpers to address these weak rotational leakage channels.

\subsection{Analysis of \texorpdfstring{$C_s$}{Cs} molecules}

\begin{table}[t]
    \centering
    \begin{tabular}{c|cc|cc|c|c|c}
              & $E$ & $E^*$ & $A'$ & $A''$ & Operators & State & $K_c$\\
              \hline
        $A'$  & 1   & 1 & $A'$ & $A''$        &$T_a,T_b,J_c$ & $\tilde{X},\tilde{A},\tilde{C}$ & even \\
        $A''$ & 1   & $-1$  & $A''$ & $A'$     &$T_c,J_a,J_b$ & $\tilde{B}$ & odd \\
    \end{tabular}
    \caption{Summary information for $C_s$, including the character table, product table, operator representations, electronic states in SrSH, and rotational state representations. Dipole operators $T_i$ and angular momentum operators $J_i$ assume the molecule is in the $ab$-plane. We also note the rotational symmetry for an $(N_{K_a K_c})$ state according to the even/odd identity of $K_c$.}
    \label{tab:Cs-summary}
\end{table}

\begin{table}[t]
    \centering
    \begin{tabular}{cccccc}
        \cline{1-5}
        \multicolumn{1}{|c|}{$A'[(0_{00}) + \epsilon(1_{10})]^+\rightarrow A'^-$} & \multicolumn{1}{c|}{$(1_{01})$} & \multicolumn{1}{c|}{$(1_{11})$} & \multicolumn{1}{c|}{$(2_{11})$} & \multicolumn{1}{c|}{$(2_{21})$} &\\
        \cline{2-5}
        \multicolumn{1}{|c|}{$A''[(0_{00})+\epsilon(1_{10})]^-\rightarrow A''^+$} & \multicolumn{1}{c|}{$a,\epsilon b$} & \multicolumn{1}{c|}{$b,\epsilon a$} & \multicolumn{1}{c|}{$\epsilon a$} & \multicolumn{1}{c|}{$\epsilon b$} &\\
        \cline{1-5}\\
        
        \hline

        \multicolumn{1}{|c|}{$A''[(0_{00}) + \epsilon(1_{10})]^-\rightarrow A'^+$} & \multicolumn{1}{c|}{$(0_{00})$} & \multicolumn{1}{c|}{$(2_{02})$} & \multicolumn{1}{c|}{$(1_{10})$} & \multicolumn{1}{c|}{$(2_{12})$} & \multicolumn{1}{c|}{$(2_{20})$} \\
        \cline{2-6}
        \multicolumn{1}{|c|}{$A'[(0_{00}) + \epsilon(1_{10})]^+\rightarrow A''^-$} & \multicolumn{1}{c|}{$\epsilon c$} & \multicolumn{1}{c|}{---} & \multicolumn{1}{c|}{$c$} & \multicolumn{1}{c|}{---} & \multicolumn{1}{c|}{$\epsilon c$} \\
        \hline
    \end{tabular}
    \caption{Possible decays from an excited state in its nominal ground rotational level, with a possible rotational state admixture, for each excited vibronic symmetry and to each ground vibronic symmetry. Superscript $\pm$ denotes parity. The second and further columns show ground rotational states and their possible population pathways. For SrSH only $A'$  vibronic levels exist in the $\X A'$ manifold but vibronic $A''$ ground levels are shown for the general case. Entries of $i=a,b,c$ denote fully allowed $i$-type transitions, while those with $\epsilon$ factors require an excited state admixture with $(1_{10})$ induced by perturbations. An entry of --- shows states compatible with $J$ and $P$ selection rules for decays, but which cannot be populated via any mechanism considered here.}
    \label{tab:Cs-decays}
\end{table}

We now consider the symmetry properties of $C_s$ molecules relevant to rotational and vibrational transitions. Summary information is presented in Tab.~\ref{tab:Cs-summary}. The operations in the $C_s$ group are $E$ (the identity) and $E^*$ (inversion, or equivalently parity). The totally symmetric representation is $A'$, and the odd-parity representation is $A''$.

In SrSH, the ground state is $\X A'$ and the first three excited states are $\A A'$, $\B A''$, and $\C A'$. All vibrational states and nuclear spin states are $A'$. Rotational states are $A'$ if $K_c$ is even, and $A''$ if $K_c$ is odd. This implies, for example, that states in the $\X A'$ manifold are even (odd) parity when $K_c$ is even (odd). By examining the character and product tables, we can see that $\X A' \rightarrow \A A'$ and $\X A' \rightarrow \C A'$ are $(a+b)$-type transitions, and $\X A' \rightarrow \B A''$ is a $c$-type transition. To a good approximation, in SrSH the $\X A' \rightarrow \A A'$ transition is $b$-type (with a small $a$-type amplitude) and the $\X A' \rightarrow \C A'$ transition is $a$-type (with a small $b$-type amplitude).

Because all vibrational states are $A'$ in SrSH, no vibrational decays are forbidden in the BOA. Here we analyze rotational decays for all possible vibronic symmetry combinations involving an upper $(0_{00};1/2)$ state. Since we assume, for purposes of this treatment, that only states with the same $J$, parity, and vibronic symmetry can mix, a $(0_{00};1/2)$ excited state in either an $A'$ or $A''$ vibronic state can mix only with a $(1_{10};1/2)$ state of the same vibronic symmetry. For example, the $(1_{11})$ rotational state (in a manifold with the same vibronic symmetry) would have opposite parity from $(0_{00})$, while a $(2_{02})$ rotational state cannot have $J=1/2$. Furthermore, since $(0_{00})$ and $(1_{10})$ have the same rotational representation, namely $A'$, there is no additional symmetry-based prohibition against these states mixing (as mentioned already in Sec.~\ref{sec:molecules}). In Table~\ref{tab:Cs-decays} we show the possible rotational states populated in a ground vibronic manifold, from a combination of $(0_{00})$ and $(1_{10})$ in an excited vibronic manifold. All ground rotational states consistent with the appropriate parity selection rule and containing a $J=1/2$ or $J=3/2$ level are presented.

We see that for $A'\rightarrow A'$ or $A''\rightarrow A''$ transitions, a nominal $(0_{00})$ state can decay to $(1_{01})$ via $a$-type transitions or to $(1_{11})$ via $b$-type transitions, as expected. However, an admixture of an excited $(1_{10})$ component can also enable decays to $(2_{11})$ ($a$-type) and $(2_{21})$ ($b$-type) as well as additional decay pathways to $(1_{01})$ ($b$-type) and $(1_{11})$ ($a$-type). For example, in SrSH even if the \Acs--\Xcs\ transition dipole were exclusively $b$-type, an excited state admixture of $(1_{01})$ would result in the population of $(1_{01})$ and $(2_{21})$.

Additionally, $J$ and parity selection rules alone would allow vibronic transitions $A''\rightarrow A'$ or $A'\rightarrow A''$ originating from an excited $(0_{00};1/2)$ state to populate $(0_{00})$, $(2_{02})$, $(1_{10})$, $(2_{12})$, or $(2_{20})$. As seen in Table~\ref{tab:Cs-decays}, the nominal $(0_{00})$ excited state leads to population of $(1_{10})$ while the possible admixture of $(1_{10})$ can produce decays to $(0_{00})$ and $(2_{20})$. However, because both excited state components have $K_c=0$, the $(2_{02})$ and $(2_{12})$ states with $K_c=2$ should not be populated upon decay even though they possess $J=3/2$ levels of the correct parity.

\subsection{Analysis of \texorpdfstring{$C_{2v}$}{C2v} molecules}

\begin{table}
    \centering
    \begin{tabular}{c|cccc|cccc|c|c|c}
              & $E$ & $(12)$ & $E^*$ & $(12)^*$ & $A_1$ & $A_2$ & $B_1$ & $B_2$ &Operators & State & $K_a K_c$\\
              \hline
        $A_1$ & $1$ & $1$ & $1$ & $1$ & $A_1$ & $A_2$ & $B_1$ & $B_2$       & $T_a$ & $\tilde{X}$, $\tilde{C}$ & $ee$ \\
        $A_2$ & $1$ & $1$ & $-1$ & $-1$ & $A_2$ & $A_1$ & $B_2$ & $B_1$        & $J_a$ &  & $eo$ \\
        $B_1$ & $1$ & $-1$ & $-1$ & $1$ & $B_1$ & $B_2$ & $A_1$ & $A_2$     & $T_c$, $J_b$ & $\tilde{B}$ & $oo$ \\
        $B_2$ & $1$ & $-1$ & $1$ & $-1$ & $B_2$ & $B_1$ & $A_2$ & $A_1$     & $T_b$, $J_c$ & $\tilde{A}$ & $oe$ \\
    \end{tabular}
    \caption{Summary information for $C_{2v}$, including the character table, product table, operator representations, electronic states in SrNH$_2$, and rotational state representations. Dipole operators $T_i$ and angular momentum operators $J_i$ assume the molecule is in the $ab$-plane and with the symmetry axis along $a$. We also note the rotational symmetry for an $(N_{K_a K_c})$ state according to the even/odd identity of $K_a$ and $K_c$.}
    \label{tab:C2v-character}
\end{table}

We now consider $C_{2v}$ molecules such as SrNH$_2$. The $C_{2v}$ group possesses four symmetry operations: the identity, $E$; permutation of identical nuclei, $(12)$; inversion, $E^*$; and permutation-inversion, $(12)^*$. Representations even under $(12)$ are written $A$, while those odd under $(12)$ are $B$. States even under $(12)^*$ obtain a subscript of $1$, while those odd under $(12)^*$ obtain a subscript of 2. The four representations defined in this way are summarized in Tab.~\ref{tab:C2v-character}.

A detailed treatment of the rovibronic structure is much more complicated in SrNH$_2$ compared to SrSH, for several reasons. In addition to the greater number of representations to consider,  vibrational states and nuclear states may not be in the totally symmetric representation. We begin with an analysis of the nuclear state symmetry, followed by an analysis of allowed rotational decays from a nominal $(0_{00};1/2)$ state (with any combination of upper and lower vibronic representations).

\begin{table}
    \centering
    \begin{tabular}{|c|c||c|c|c|c|}
    \hline
        $\Gamma_{ev}$ & $I_H$ & $\Gamma_{\rm{rot}}=A_1$ & $\Gamma_{\rm{rot}}=A_2$ & $\Gamma_{\rm{rot}}=B_1$ & $\Gamma_{\rm{rot}}=B_2$\\
        \hline\hline
        $A_1$ & $1/2$ & $B_2,0;B_2,+$ & $B_2,0;B_1,-$ & $A_1,1;B_1,-$ & $A_1,1;B_2,+$\\
        $A_2$ & $1/2$ & $B_2,0;B_1,-$ & $B_2,0;B_2,+$ & $A_1,1;B_2,+$ & $A_1,1;B_1,-$\\
        $B_1$ & $1/2$ & $A_1,1;B_1,-$ & $A_1,1;B_2,+$ & $B_2,0;B_2,+$ & $B_2,0;B_1,-$\\
        $B_2$ & $1/2$ & $A_1,1;B_2,+$ & $A_1,1;B_1,-$ & $B_2,0;B_1,-$ & $B_2,0;B_2,+$\\
        \hline
        $A_1$ & $1$   & $A_1,e;A_1,+$ & $A_1,e;A_2,-$ & $B_2,1;A_2,-$ & $B_2,1;A_1,+$\\
        $A_2$ & $1$   & $A_1,e;A_2,-$ & $A_1,e;A_1,+$ & $B_2,1;A_1,+$ & $B_2,1;A_2,-$\\
        $B_1$ & $1$   & $B_2,1;A_2,-$ & $B_2,1;A_1,+$ & $A_1,e;A_1,+$ & $A_1,e;A_2,-$\\
        $B_2$ & $1$   & $B_2,1;A_1,+$ & $B_2,1;A_2,-$ & $A_1,e;A_2,-$ & $A_1,e;A_1,+$\\
        \hline
    \end{tabular}
    \caption{Relationship between vibronic representation, nuclear spin, and rotational representation in $C_{2v}$ molecules. Rows correspond to vibronic representations $\Gamma_{ev}$ and hydrogen nuclear spins $I_H$, while columns correspond to rotational representations. Entries are of the form $\Gamma_{\text{nuc}},I;\Gamma_{\text{tot}},P$ specifying the required nuclear state representation, nuclear spin, total molecular state representation, and associated state parity. A nuclear spin of $e$ indicates allowed values of $I=0$ or $I=2$.}
    \label{tab:C2v-rovibronic}
\end{table}

Directly applying symmetry operations to nuclear states allows a given state's representation to be determined. For ordinary hydrogen with $I_H=1/2$, the triplet $I=1$ state has symmetry $A_1$ and the singlet $I=0$ state has symmetry $B_2$. For deuterium with $I_H=1$, the quintet with $I=2$ and singlet with $I=0$ both have representation $A_1$ and the triplet with $I=1$ has representation $B_2$. With identical fermions, the total molecular wave function must be odd under hydrogen exchange and therefore transform as $B_1$ or $B_2$. On the other hand, with identical bosons, the total molecular wave function must be even under hydrogen (i.e., deuterium) exchange and therefore transform as $A_1$ or $A_2$. This implies that not all nuclear spin states are compatible with a given rovibronic state. The allowed nuclear state representation, nuclear spin states, total molecular representation, and total state parity for every combination of vibronic representation, rotational representation, and hydrogen spin magnitude is shown in Table~\ref{tab:C2v-rovibronic}.

The rotational representation of an $(N_{K_a K_c})$ state depends on whether both $K_a$ and $K_c$ are even or odd. However, the parity of a state is, as for $C_s$, determined only by whether $K_c$ is even or odd (see Table~\ref{tab:C2v-character}). Thus the combination of parity and $J$ selection rules alone would enable $(0_{00})$ in an excited vibronic manifold to mix with, at most, a $(1_{10})$ state with the same vibronic symmetry. However, unlike in the $C_s$ case, the $(0_{00})$ and $(1_{10})$ states have different rotational symmetries and therefore do not ordinarily mix under the classes of perturbations we consider, described in Sec.~\ref{sec:transition-rules}. Specifically, $(0_{00})$ has $A_1$ symmetry while $(1_{10})$ has $B_2$ symmetry. Viewed another way, as seen in Table~\ref{tab:C2v-rovibronic}, the $(0_{00})$ and $(1_{10})$ states exist for different nuclear isomers and can only mix via hyperfine interactions, provided the vibronic representation is pure.

We show the resulting rovibronic transitions in Table~\ref{tab:C2v-decays}. For completeness, we construct tables with the possibility of mixing excited state $(0_{00})$ and $(1_{00})$ components, but adopt a new rotational state label notation $(N_{K_a K_c}^I)$ to emphasize the appropriate nuclear state $I$ in each case. As already mentioned, in every case the admixture of $(1_{10})$ into $(0_{00})$ requires mixing states with different total nuclear spin values. If we assume that such a mixing is absolutely negligible, $\epsilon\rightarrow0$, then for every combination of ground and excited vibronic representations there is exactly one rotational state that can be populated from the excited $(0_{00};1/2)$ level (except for cases where the vibronic manifolds differ by an $A_2$ representation, in which case the transition is vibronically forbidden regardless of the rotational levels involved). These results do not depend on the value of the hydrogen spin, $I_H$. The increased symmetry of $C_{2v}$ compared to $C_s$ therefore provides strong protection against rotational leakage channels.

\begin{table}
    \centering
    \bgroup
    \def\arraystretch{1.5}
    \begin{tabular}{cccccc}
        
        \cline{1-5}
        \multicolumn{1}{|c|}{$A_1[(0_{00}^e)+\epsilon(1_{10}^1)]^+\rightarrow A_1^-$} & \multicolumn{1}{c|}{$(1_{01}^e)$} & \multicolumn{1}{c|}{$(2_{21}^e)$} & \multicolumn{1}{c|}{$(1_{11}^1)$} & \multicolumn{1}{c|}{$(2_{11}^1)$} &\\
        \cline{2-5}
        \multicolumn{1}{|c|}{$A_2[(0_{00}^e)+\epsilon(1_{10}^1)]^-\rightarrow A_2^+$} & \multicolumn{1}{c|}{$a$} &  \multicolumn{1}{c|}{---} & \multicolumn{1}{c|}{$\epsilon a$} & \multicolumn{1}{c|}{$\epsilon a$}  &\\
        
        \hhline{=====~}
        
        \multicolumn{1}{|c|}{$B_1[(0_{00}^1)+\epsilon(1_{10}^e)]^-\rightarrow B_1^+$} & \multicolumn{1}{c|}{$(1_{01}^1)$} & \multicolumn{1}{c|}{$(2_{21}^1)$} & \multicolumn{1}{c|}{$(1_{11}^e)$} & \multicolumn{1}{c|}{$(2_{11}^e)$} &\\
        \cline{2-5}
        \multicolumn{1}{|c|}{$B_2[(0_{00}^1)+\epsilon(1_{10}^e)]^+\rightarrow B_2^-$} & \multicolumn{1}{c|}{$a$} & \multicolumn{1}{c|}{---}  & \multicolumn{1}{c|}{$\epsilon a$}  &  \multicolumn{1}{c|}{$\epsilon a$}  &\\

        \hhline{=====~}

        \multicolumn{1}{|c|}{$A_1[(0_{00}^e)+\epsilon(1_{10}^1)]^+\rightarrow B_2^-$} & \multicolumn{1}{c|}{$(1_{01}^1)$} & \multicolumn{1}{c|}{$(2_{21}^1)$} & \multicolumn{1}{c|}{$(1_{11}^e)$} & \multicolumn{1}{c|}{$(2_{11}^e)$} &\\
        \cline{2-5}
        \multicolumn{1}{|c|}{$A_2[(0_{00}^e)+\epsilon(1_{10}^1)]^-\rightarrow B_1^+$} & \multicolumn{1}{c|}{$\epsilon b$} & \multicolumn{1}{c|}{$\epsilon b$}  & \multicolumn{1}{c|}{$b$}  & \multicolumn{1}{c|}{---}  &\\
        
        \hhline{=====~}
        \multicolumn{1}{|c|}{$B_1[(0_{00}^1)+\epsilon(1_{10}^e)]^-\rightarrow A_2^+$} & \multicolumn{1}{c|}{$(1_{01}^e)$} & \multicolumn{1}{c|}{$(2_{21}^e)$} & \multicolumn{1}{c|}{$(1_{11}^1)$} & \multicolumn{1}{c|}{$(2_{11}^1)$} &\\
        \cline{2-5}
        \multicolumn{1}{|c|}{$B_2[(0_{00}^1)+\epsilon(1_{10}^e)]^+\rightarrow A_1^-$} & \multicolumn{1}{c|}{$\epsilon b$} & \multicolumn{1}{c|}{$\epsilon b$}  & \multicolumn{1}{c|}{$b$}  & \multicolumn{1}{c|}{---}  &\\
        \cline{1-5} \\
        
        \hline
        \multicolumn{1}{|c|}{$A_1[(0_{00}^e)+\epsilon(1_{10}^1)]^+\rightarrow B_1^-$} & \multicolumn{1}{c|}{$(0_{00}^1)$} & \multicolumn{1}{c|}{$(2_{02}^1)$} & \multicolumn{1}{c|}{$(2_{20}^1)$} & \multicolumn{1}{c|}{$(1_{10}^e)$} & \multicolumn{1}{c|}{$(2_{12}^e)$} \\
        \cline{2-6}
        \multicolumn{1}{|c|}{$A_2[(0_{00}^e)+\epsilon(1_{10}^1)]^-\rightarrow B_2^+$} & \multicolumn{1}{c|}{$\epsilon c$} & \multicolumn{1}{c|}{---}  & \multicolumn{1}{c|}{$\epsilon c$}  & \multicolumn{1}{c|}{$c$}  & \multicolumn{1}{c|}{---}\\
        \hline
        \hline
        
        \multicolumn{1}{|c|}{$B_1[(0_{00}^1)+\epsilon(1_{10}^e)]^-\rightarrow A_1^+$} & \multicolumn{1}{c|}{$(0_{00}^e)$} & \multicolumn{1}{c|}{$(2_{02}^e)$} & \multicolumn{1}{c|}{$(2_{20}^e)$} & \multicolumn{1}{c|}{$(1_{10}^1)$} & \multicolumn{1}{c|}{$(2_{12}^1)$} \\
        \cline{2-6}
        \multicolumn{1}{|c|}{$B_2[(0_{00}^1)+\epsilon(1_{10}^e)]^+\rightarrow A_2^-$} & \multicolumn{1}{c|}{$\epsilon c$} & \multicolumn{1}{c|}{---}  &  \multicolumn{1}{c|}{$\epsilon c$} & \multicolumn{1}{c|}{$c$}  & \multicolumn{1}{c|}{---}\\
        \hline
        \hline

        \multicolumn{1}{|c|}{$A_1[(0_{00}^e)+\epsilon(1_{10}^1)]^+\rightarrow A_2^-$} & \multicolumn{1}{c|}{$(0_{00}^e)$} & \multicolumn{1}{c|}{$(2_{02}^e)$} & \multicolumn{1}{c|}{$(2_{20}^e)$} & \multicolumn{1}{c|}{$(1_{10}^1)$} & \multicolumn{1}{c|}{$(2_{12}^1)$} \\
        \cline{2-6}
        \multicolumn{1}{|c|}{$A_2[(0_{00}^e)+\epsilon(1_{10}^1)]^-\rightarrow A_1^+$} & \multicolumn{1}{c|}{---} & \multicolumn{1}{c|}{---}  & \multicolumn{1}{c|}{---}  & \multicolumn{1}{c|}{---}  & \multicolumn{1}{c|}{---}\\
        \hline
        \hline
        
        \multicolumn{1}{|c|}{$B_1[(0_{00}^1)+\epsilon(1_{10}^e)]^-\rightarrow B_2^+$} & \multicolumn{1}{c|}{$(0_{00}^1)$} & \multicolumn{1}{c|}{$(2_{02}^1)$} & \multicolumn{1}{c|}{$(2_{20}^1)$} & \multicolumn{1}{c|}{$(1_{10}^e)$} & \multicolumn{1}{c|}{$(2_{12}^e)$} \\
        \cline{2-6}
        \multicolumn{1}{|c|}{$B_2[(0_{00}^1)+\epsilon(1_{10}^e)]^+\rightarrow B_1^-$} & \multicolumn{1}{c|}{---} & \multicolumn{1}{c|}{---}  &  \multicolumn{1}{c|}{---} & \multicolumn{1}{c|}{---}  & \multicolumn{1}{c|}{---}\\
        \hline
    \end{tabular}
    \egroup
    \caption{Allowed decays from an excited state in its nominal ground rotational level, with possible rotational state admixture of $(1_{10})$, for each excited vibronic representation and to each ground vibronic representation. Superscript $\pm$ denotes parity, and superscripts within a rotational state label denotes nuclear spin $I$. For $I_H=1/2$, $e$ signifies $I=0$ while for $I_H=1$, $e$ signifies $I=0$ or $2$. The second and further columns show ground rotational states and their possible population pathways. Entries of $i=a,b,c$ denote fully allowed $i$-type transitions, while those with $\epsilon$ factors require an excited state admixture with $(1_{10})$ induced by perturbations. Entries of --- show states compatible with $J$ and $P$ selection rules on decays, but which cannot be populated via mechanisms considered here.}
    \label{tab:C2v-decays}
\end{table}

\subsection{Analysis of \texorpdfstring{$C_1$}{C1} molecules}

Totally asymmetric (chiral) molecules such as SrOCHDT have also been considered for laser cooling. In this case, the only operation in the molecular symmetry group is the identity, $E$, and the only representation is the totally symmetric representation, $A$. A molecule in $C_1$ cannot be in a parity eigenstate because parity maps an enantiomer to its opposite chirality, regardless of the rotational composition. In principle, one may obtain parity eigenstates of a chiral molecule by considering an extended set of states that includes both left-handed and right-handed configurations; the parity eigenstates are constructed from even and odd linear combinations of enantiomers. However, in most practical cases the tunneling time between enantiomers is extremely long and a chiral isomer may be considered in isolation, so that the $C_1$ molecular symmetry group is appropriate.

In this situation, rotational branching is constrained only by $J$ selection rules. An excited $(0_{00})$ state can decay to the $J=1/2$ or $J=3/2$ sublevels of $(1_{10})$, $(1_{01})$, or $(1_{11})$. Furthermore, excited state rotational mixing of $(0_{00})$ with $(1_{01})$, $(1_{10})$, and $(1_{11})$ may enable additional decays to $(0_{00})$, $(2_{02})$, $(2_{11})$, $(2_{12})$, $(2_{20})$, and $(2_{21})$. Nevertheless, the magnitudes of any excited state mixing or transition strength must be assessed on a case-by-case basis, and these 9 ground rotational states will not typically be comparably populated.

\bibliographystyle{apsrev4-2}
\bibliography{VBRbibv2}

\begin{thebibliography}{85}%
\makeatletter
\providecommand \@ifxundefined [1]{%
 \@ifx{#1\undefined}
}%
\providecommand \@ifnum [1]{%
 \ifnum #1\expandafter \@firstoftwo
 \else \expandafter \@secondoftwo
 \fi
}%
\providecommand \@ifx [1]{%
 \ifx #1\expandafter \@firstoftwo
 \else \expandafter \@secondoftwo
 \fi
}%
\providecommand \natexlab [1]{#1}%
\providecommand \enquote  [1]{``#1''}%
\providecommand \bibnamefont  [1]{#1}%
\providecommand \bibfnamefont [1]{#1}%
\providecommand \citenamefont [1]{#1}%
\providecommand \href@noop [0]{\@secondoftwo}%
\providecommand \href [0]{\begingroup \@sanitize@url \@href}%
\providecommand \@href[1]{\@@startlink{#1}\@@href}%
\providecommand \@@href[1]{\endgroup#1\@@endlink}%
\providecommand \@sanitize@url [0]{\catcode `\\12\catcode `\$12\catcode `\&12\catcode `\#12\catcode `\^12\catcode `\_12\catcode `\%12\relax}%
\providecommand \@@startlink[1]{}%
\providecommand \@@endlink[0]{}%
\providecommand \url  [0]{\begingroup\@sanitize@url \@url }%
\providecommand \@url [1]{\endgroup\@href {#1}{\urlprefix }}%
\providecommand \urlprefix  [0]{URL }%
\providecommand \Eprint [0]{\href }%
\providecommand \doibase [0]{https://doi.org/}%
\providecommand \selectlanguage [0]{\@gobble}%
\providecommand \bibinfo  [0]{\@secondoftwo}%
\providecommand \bibfield  [0]{\@secondoftwo}%
\providecommand \translation [1]{[#1]}%
\providecommand \BibitemOpen [0]{}%
\providecommand \bibitemStop [0]{}%
\providecommand \bibitemNoStop [0]{.\EOS\space}%
\providecommand \EOS [0]{\spacefactor3000\relax}%
\providecommand \BibitemShut  [1]{\csname bibitem#1\endcsname}%
\let\auto@bib@innerbib\@empty
\bibitem [{\citenamefont {Hao}\ \emph {et~al.}(2020)\citenamefont {Hao}, \citenamefont {Navr\'atil}, \citenamefont {Norrgard}, \citenamefont {Ilia\ifmmode~\check{s}\else \v{s}\fi{}}, \citenamefont {Eliav}, \citenamefont {Timmermans}, \citenamefont {Flambaum},\ and\ \citenamefont {Borschevsky}}]{Hao2020}%
  \BibitemOpen
  \bibfield  {author} {\bibinfo {author} {\bibfnamefont {Y.}~\bibnamefont {Hao}}, \bibinfo {author} {\bibfnamefont {P.}~\bibnamefont {Navr\'atil}}, \bibinfo {author} {\bibfnamefont {E.~B.}\ \bibnamefont {Norrgard}}, \bibinfo {author} {\bibfnamefont {M.}~\bibnamefont {Ilia\ifmmode~\check{s}\else \v{s}\fi{}}}, \bibinfo {author} {\bibfnamefont {E.}~\bibnamefont {Eliav}}, \bibinfo {author} {\bibfnamefont {R.~G.~E.}\ \bibnamefont {Timmermans}}, \bibinfo {author} {\bibfnamefont {V.~V.}\ \bibnamefont {Flambaum}},\ and\ \bibinfo {author} {\bibfnamefont {A.}~\bibnamefont {Borschevsky}},\ }\href {https://link.aps.org/doi/10.1103/PhysRevA.102.052828} {\bibfield  {journal} {\bibinfo  {journal} {Phys. Rev. A}\ }\textbf {\bibinfo {volume} {102}},\ \bibinfo {pages} {052828} (\bibinfo {year} {2020})}\BibitemShut {NoStop}%
\bibitem [{\citenamefont {Kozyryev}\ \emph {et~al.}(2021)\citenamefont {Kozyryev}, \citenamefont {Lasner},\ and\ \citenamefont {Doyle}}]{Kozyryev2021}%
  \BibitemOpen
  \bibfield  {author} {\bibinfo {author} {\bibfnamefont {I.}~\bibnamefont {Kozyryev}}, \bibinfo {author} {\bibfnamefont {Z.}~\bibnamefont {Lasner}},\ and\ \bibinfo {author} {\bibfnamefont {J.~M.}\ \bibnamefont {Doyle}},\ }\href {https://journals.aps.org/pra/abstract/10.1103/PhysRevA.103.043313} {\bibfield  {journal} {\bibinfo  {journal} {Phys. Rev. A}\ }\textbf {\bibinfo {volume} {103}},\ \bibinfo {pages} {043313} (\bibinfo {year} {2021})}\BibitemShut {NoStop}%
\bibitem [{\citenamefont {Isaev}\ and\ \citenamefont {Berger}(2018)}]{Isaev2018}%
  \BibitemOpen
  \bibfield  {author} {\bibinfo {author} {\bibfnamefont {T.~A.}\ \bibnamefont {Isaev}}\ and\ \bibinfo {author} {\bibfnamefont {R.}~\bibnamefont {Berger}},\ }\href {https://www.chimia.ch/chimia/article/view/1445} {\bibfield  {journal} {\bibinfo  {journal} {CHIMIA}\ }\textbf {\bibinfo {volume} {72}},\ \bibinfo {pages} {375} (\bibinfo {year} {2018})}\BibitemShut {NoStop}%
\bibitem [{\citenamefont {Norrgard}\ \emph {et~al.}(2019)\citenamefont {Norrgard}, \citenamefont {Barker}, \citenamefont {Eckel}, \citenamefont {Fedchak}, \citenamefont {Klimov},\ and\ \citenamefont {Scherschligt}}]{Norrgard2019}%
  \BibitemOpen
  \bibfield  {author} {\bibinfo {author} {\bibfnamefont {E.~B.}\ \bibnamefont {Norrgard}}, \bibinfo {author} {\bibfnamefont {D.~S.}\ \bibnamefont {Barker}}, \bibinfo {author} {\bibfnamefont {S.}~\bibnamefont {Eckel}}, \bibinfo {author} {\bibfnamefont {J.~A.}\ \bibnamefont {Fedchak}}, \bibinfo {author} {\bibfnamefont {N.~N.}\ \bibnamefont {Klimov}},\ and\ \bibinfo {author} {\bibfnamefont {J.}~\bibnamefont {Scherschligt}},\ }\href {https://www.nature.com/articles/s42005-019-0181-1} {\bibfield  {journal} {\bibinfo  {journal} {Commun. Phys.}\ }\textbf {\bibinfo {volume} {2}},\ \bibinfo {pages} {77} (\bibinfo {year} {2019})}\BibitemShut {NoStop}%
\bibitem [{\citenamefont {Kozyryev}\ and\ \citenamefont {Hutzler}(2017)}]{Kozyryev2017}%
  \BibitemOpen
  \bibfield  {author} {\bibinfo {author} {\bibfnamefont {I.}~\bibnamefont {Kozyryev}}\ and\ \bibinfo {author} {\bibfnamefont {N.~R.}\ \bibnamefont {Hutzler}},\ }\href {https://journals.aps.org/prl/abstract/10.1103/PhysRevLett.119.133002} {\bibfield  {journal} {\bibinfo  {journal} {Phys. Rev. Lett.}\ }\textbf {\bibinfo {volume} {119}},\ \bibinfo {pages} {133002} (\bibinfo {year} {2017})}\BibitemShut {NoStop}%
\bibitem [{\citenamefont {Isaev}\ \emph {et~al.}(2017)\citenamefont {Isaev}, \citenamefont {Zaitsevskii},\ and\ \citenamefont {Eliav}}]{Isaev2017}%
  \BibitemOpen
  \bibfield  {author} {\bibinfo {author} {\bibfnamefont {T.~A.}\ \bibnamefont {Isaev}}, \bibinfo {author} {\bibfnamefont {A.~V.}\ \bibnamefont {Zaitsevskii}},\ and\ \bibinfo {author} {\bibfnamefont {E.}~\bibnamefont {Eliav}},\ }\href {https://www.doi.org/10.1088/1361-6455/aa8f34} {\bibfield  {journal} {\bibinfo  {journal} {J. Phys. B}\ }\textbf {\bibinfo {volume} {50}},\ \bibinfo {pages} {225101} (\bibinfo {year} {2017})}\BibitemShut {NoStop}%
\bibitem [{\citenamefont {Denis}\ \emph {et~al.}(2020)\citenamefont {Denis}, \citenamefont {Hao}, \citenamefont {Eliav}, \citenamefont {Hutzler}, \citenamefont {Nayak}, \citenamefont {Timmermans},\ and\ \citenamefont {Borschesvky}}]{Denis2020}%
  \BibitemOpen
  \bibfield  {author} {\bibinfo {author} {\bibfnamefont {M.}~\bibnamefont {Denis}}, \bibinfo {author} {\bibfnamefont {Y.}~\bibnamefont {Hao}}, \bibinfo {author} {\bibfnamefont {E.}~\bibnamefont {Eliav}}, \bibinfo {author} {\bibfnamefont {N.~R.}\ \bibnamefont {Hutzler}}, \bibinfo {author} {\bibfnamefont {M.~K.}\ \bibnamefont {Nayak}}, \bibinfo {author} {\bibfnamefont {R.~G.}\ \bibnamefont {Timmermans}},\ and\ \bibinfo {author} {\bibfnamefont {A.}~\bibnamefont {Borschesvky}},\ }\href {https://doi.org/10.1063/1.5141065} {\bibfield  {journal} {\bibinfo  {journal} {J. Chem. Phys.}\ }\textbf {\bibinfo {volume} {152}},\ \bibinfo {pages} {84303} (\bibinfo {year} {2020})}\BibitemShut {NoStop}%
\bibitem [{\citenamefont {Hutzler}(2020)}]{Hutzler2020}%
  \BibitemOpen
  \bibfield  {author} {\bibinfo {author} {\bibfnamefont {N.~R.}\ \bibnamefont {Hutzler}},\ }\href {https://iopscience.iop.org/article/10.1088/2058-9565/abb9c5} {\bibfield  {journal} {\bibinfo  {journal} {Quant. Sci. Tech.}\ }\textbf {\bibinfo {volume} {5}},\ \bibinfo {pages} {044011} (\bibinfo {year} {2020})}\BibitemShut {NoStop}%
\bibitem [{\citenamefont {Safronova}\ \emph {et~al.}(2018)\citenamefont {Safronova}, \citenamefont {Budker}, \citenamefont {DeMille}, \citenamefont {Kimball}, \citenamefont {Derevianko},\ and\ \citenamefont {Clark}}]{Safronova2018}%
  \BibitemOpen
  \bibfield  {author} {\bibinfo {author} {\bibfnamefont {M.~S.}\ \bibnamefont {Safronova}}, \bibinfo {author} {\bibfnamefont {D.}~\bibnamefont {Budker}}, \bibinfo {author} {\bibfnamefont {D.}~\bibnamefont {DeMille}}, \bibinfo {author} {\bibfnamefont {D.~F.~J.}\ \bibnamefont {Kimball}}, \bibinfo {author} {\bibfnamefont {A.}~\bibnamefont {Derevianko}},\ and\ \bibinfo {author} {\bibfnamefont {C.~W.}\ \bibnamefont {Clark}},\ }\href {https://link.aps.org/doi/10.1103/RevModPhys.90.025008} {\bibfield  {journal} {\bibinfo  {journal} {Rev. Mod. Phys.}\ }\textbf {\bibinfo {volume} {90}},\ \bibinfo {pages} {025008} (\bibinfo {year} {2018})}\BibitemShut {NoStop}%
\bibitem [{\citenamefont {Eckel}\ \emph {et~al.}(2013)\citenamefont {Eckel}, \citenamefont {Hamilton}, \citenamefont {Kirilov}, \citenamefont {Smith},\ and\ \citenamefont {DeMille}}]{Eckel2013}%
  \BibitemOpen
  \bibfield  {author} {\bibinfo {author} {\bibfnamefont {S.}~\bibnamefont {Eckel}}, \bibinfo {author} {\bibfnamefont {P.}~\bibnamefont {Hamilton}}, \bibinfo {author} {\bibfnamefont {E.}~\bibnamefont {Kirilov}}, \bibinfo {author} {\bibfnamefont {H.~W.}\ \bibnamefont {Smith}},\ and\ \bibinfo {author} {\bibfnamefont {D.}~\bibnamefont {DeMille}},\ }\href {https://link.aps.org/doi/10.1103/PhysRevA.87.052130} {\bibfield  {journal} {\bibinfo  {journal} {Phys. Rev. A}\ }\textbf {\bibinfo {volume} {87}},\ \bibinfo {pages} {052130} (\bibinfo {year} {2013})}\BibitemShut {NoStop}%
\bibitem [{\citenamefont {DeMille}(2015)}]{DeMille2015}%
  \BibitemOpen
  \bibfield  {author} {\bibinfo {author} {\bibfnamefont {D.}~\bibnamefont {DeMille}},\ }\href {https://pubs.aip.org/physicstoday/article/68/12/34/415072/Diatomic-molecules-a-window-onto-fundamental} {\bibfield  {journal} {\bibinfo  {journal} {Physics Today}\ }\textbf {\bibinfo {volume} {68}},\ \bibinfo {pages} {34} (\bibinfo {year} {2015})}\BibitemShut {NoStop}%
\bibitem [{\citenamefont {Anderegg}\ \emph {et~al.}(2017)\citenamefont {Anderegg}, \citenamefont {Augenbraun}, \citenamefont {Chae}, \citenamefont {Hemmerling}, \citenamefont {Hutzler}, \citenamefont {Ravi}, \citenamefont {Collopy}, \citenamefont {Ye}, \citenamefont {Ketterle},\ and\ \citenamefont {Doyle}}]{Anderegg2017}%
  \BibitemOpen
  \bibfield  {author} {\bibinfo {author} {\bibfnamefont {L.}~\bibnamefont {Anderegg}}, \bibinfo {author} {\bibfnamefont {B.~L.}\ \bibnamefont {Augenbraun}}, \bibinfo {author} {\bibfnamefont {E.}~\bibnamefont {Chae}}, \bibinfo {author} {\bibfnamefont {B.}~\bibnamefont {Hemmerling}}, \bibinfo {author} {\bibfnamefont {N.~R.}\ \bibnamefont {Hutzler}}, \bibinfo {author} {\bibfnamefont {A.}~\bibnamefont {Ravi}}, \bibinfo {author} {\bibfnamefont {A.}~\bibnamefont {Collopy}}, \bibinfo {author} {\bibfnamefont {J.}~\bibnamefont {Ye}}, \bibinfo {author} {\bibfnamefont {W.}~\bibnamefont {Ketterle}},\ and\ \bibinfo {author} {\bibfnamefont {J.~M.}\ \bibnamefont {Doyle}},\ }\href {https://journals.aps.org/prl/abstract/10.1103/PhysRevLett.119.103201} {\bibfield  {journal} {\bibinfo  {journal} {Phys. Rev. Lett.}\ }\textbf {\bibinfo {volume} {119}},\ \bibinfo {pages} {103201} (\bibinfo {year} {2017})}\BibitemShut {NoStop}%
\bibitem [{\citenamefont {V{\'{a}}zquez-Carson}\ \emph {et~al.}(2022)\citenamefont {V{\'{a}}zquez-Carson}, \citenamefont {Sun}, \citenamefont {Dai}, \citenamefont {Mitra},\ and\ \citenamefont {Zelevinsky}}]{Vazquez-Carson2022}%
  \BibitemOpen
  \bibfield  {author} {\bibinfo {author} {\bibfnamefont {S.~F.}\ \bibnamefont {V{\'{a}}zquez-Carson}}, \bibinfo {author} {\bibfnamefont {Q.}~\bibnamefont {Sun}}, \bibinfo {author} {\bibfnamefont {J.}~\bibnamefont {Dai}}, \bibinfo {author} {\bibfnamefont {D.}~\bibnamefont {Mitra}},\ and\ \bibinfo {author} {\bibfnamefont {T.}~\bibnamefont {Zelevinsky}},\ }\href {https://iopscience.iop.org/article/10.1088/1367-2630/ac806c/meta} {\bibfield  {journal} {\bibinfo  {journal} {New J. Phys.}\ }\textbf {\bibinfo {volume} {24}},\ \bibinfo {pages} {083006} (\bibinfo {year} {2022})}\BibitemShut {NoStop}%
\bibitem [{\citenamefont {Barry}\ \emph {et~al.}(2014)\citenamefont {Barry}, \citenamefont {McCarron}, \citenamefont {Norrgard}, \citenamefont {Steinecker},\ and\ \citenamefont {Demille}}]{Barry2014}%
  \BibitemOpen
  \bibfield  {author} {\bibinfo {author} {\bibfnamefont {J.~F.}\ \bibnamefont {Barry}}, \bibinfo {author} {\bibfnamefont {D.~J.}\ \bibnamefont {McCarron}}, \bibinfo {author} {\bibfnamefont {E.~B.}\ \bibnamefont {Norrgard}}, \bibinfo {author} {\bibfnamefont {M.~H.}\ \bibnamefont {Steinecker}},\ and\ \bibinfo {author} {\bibfnamefont {D.}~\bibnamefont {Demille}},\ }\href {https://www.nature.com/articles/nature13634} {\bibfield  {journal} {\bibinfo  {journal} {Nature}\ }\textbf {\bibinfo {volume} {512}},\ \bibinfo {pages} {286} (\bibinfo {year} {2014})}\BibitemShut {NoStop}%
\bibitem [{\citenamefont {Collopy}\ \emph {et~al.}(2018)\citenamefont {Collopy}, \citenamefont {Ding}, \citenamefont {Wu}, \citenamefont {Finneran}, \citenamefont {Anderegg}, \citenamefont {Augenbraun}, \citenamefont {Doyle},\ and\ \citenamefont {Ye}}]{Collopy2018}%
  \BibitemOpen
  \bibfield  {author} {\bibinfo {author} {\bibfnamefont {A.~L.}\ \bibnamefont {Collopy}}, \bibinfo {author} {\bibfnamefont {S.}~\bibnamefont {Ding}}, \bibinfo {author} {\bibfnamefont {Y.}~\bibnamefont {Wu}}, \bibinfo {author} {\bibfnamefont {I.~A.}\ \bibnamefont {Finneran}}, \bibinfo {author} {\bibfnamefont {L.}~\bibnamefont {Anderegg}}, \bibinfo {author} {\bibfnamefont {B.~L.}\ \bibnamefont {Augenbraun}}, \bibinfo {author} {\bibfnamefont {J.~M.}\ \bibnamefont {Doyle}},\ and\ \bibinfo {author} {\bibfnamefont {J.}~\bibnamefont {Ye}},\ }\href {https://journals.aps.org/prl/abstract/10.1103/PhysRevLett.121.213201} {\bibfield  {journal} {\bibinfo  {journal} {Phys. Rev. Lett.}\ }\textbf {\bibinfo {volume} {121}},\ \bibinfo {pages} {213201} (\bibinfo {year} {2018})}\BibitemShut {NoStop}%
\bibitem [{\citenamefont {Alauze}\ \emph {et~al.}(2021)\citenamefont {Alauze}, \citenamefont {Lim}, \citenamefont {Trigatzis}, \citenamefont {Swarbrick}, \citenamefont {Collings}, \citenamefont {Fitch}, \citenamefont {Sauer},\ and\ \citenamefont {Tarbutt}}]{Alauze2021}%
  \BibitemOpen
  \bibfield  {author} {\bibinfo {author} {\bibfnamefont {X.}~\bibnamefont {Alauze}}, \bibinfo {author} {\bibfnamefont {J.}~\bibnamefont {Lim}}, \bibinfo {author} {\bibfnamefont {M.~A.}\ \bibnamefont {Trigatzis}}, \bibinfo {author} {\bibfnamefont {S.}~\bibnamefont {Swarbrick}}, \bibinfo {author} {\bibfnamefont {F.~J.}\ \bibnamefont {Collings}}, \bibinfo {author} {\bibfnamefont {N.~J.}\ \bibnamefont {Fitch}}, \bibinfo {author} {\bibfnamefont {B.~E.}\ \bibnamefont {Sauer}},\ and\ \bibinfo {author} {\bibfnamefont {M.~R.}\ \bibnamefont {Tarbutt}},\ }\href {https://iopscience.iop.org/article/10.1088/2058-9565/ac107e/meta} {\bibfield  {journal} {\bibinfo  {journal} {Quant. Sci. Tech.}\ }\textbf {\bibinfo {volume} {6}},\ \bibinfo {pages} {044005} (\bibinfo {year} {2021})}\BibitemShut {NoStop}%
\bibitem [{\citenamefont {Zeng}\ \emph {et~al.}(2024)\citenamefont {Zeng}, \citenamefont {Deng}, \citenamefont {Yang},\ and\ \citenamefont {Yan}}]{Zeng2024}%
  \BibitemOpen
  \bibfield  {author} {\bibinfo {author} {\bibfnamefont {Z.}~\bibnamefont {Zeng}}, \bibinfo {author} {\bibfnamefont {S.}~\bibnamefont {Deng}}, \bibinfo {author} {\bibfnamefont {S.}~\bibnamefont {Yang}},\ and\ \bibinfo {author} {\bibfnamefont {B.}~\bibnamefont {Yan}},\ }\href {https://arxiv.org/abs/2405.17883v1} {\  (\bibinfo {year} {2024})},\ \Eprint {https://arxiv.org/abs/2405.17883} {arXiv:2405.17883} \BibitemShut {NoStop}%
\bibitem [{\citenamefont {McNally}\ \emph {et~al.}(2020)\citenamefont {McNally}, \citenamefont {Kozyryev}, \citenamefont {Vasquez-Carson}, \citenamefont {Wenz}, \citenamefont {Wang},\ and\ \citenamefont {Zelevinsky}}]{mcnally2020optical}%
  \BibitemOpen
  \bibfield  {author} {\bibinfo {author} {\bibfnamefont {R.~L.}\ \bibnamefont {McNally}}, \bibinfo {author} {\bibfnamefont {I.}~\bibnamefont {Kozyryev}}, \bibinfo {author} {\bibfnamefont {S.}~\bibnamefont {Vasquez-Carson}}, \bibinfo {author} {\bibfnamefont {K.}~\bibnamefont {Wenz}}, \bibinfo {author} {\bibfnamefont {T.}~\bibnamefont {Wang}},\ and\ \bibinfo {author} {\bibfnamefont {T.}~\bibnamefont {Zelevinsky}},\ }\href {https://doi.org/10.1088/1367-2630/aba3e9} {\bibfield  {journal} {\bibinfo  {journal} {New J. Phys.}\ }\textbf {\bibinfo {volume} {22}},\ \bibinfo {pages} {083047} (\bibinfo {year} {2020})}\BibitemShut {NoStop}%
\bibitem [{\citenamefont {Hallas}\ \emph {et~al.}(2023)\citenamefont {Hallas}, \citenamefont {Vilas}, \citenamefont {Anderegg}, \citenamefont {Robichaud}, \citenamefont {Winnicki}, \citenamefont {Zhang}, \citenamefont {Cheng},\ and\ \citenamefont {Doyle}}]{hallas2023caohtrap}%
  \BibitemOpen
  \bibfield  {author} {\bibinfo {author} {\bibfnamefont {C.}~\bibnamefont {Hallas}}, \bibinfo {author} {\bibfnamefont {N.~B.}\ \bibnamefont {Vilas}}, \bibinfo {author} {\bibfnamefont {L.}~\bibnamefont {Anderegg}}, \bibinfo {author} {\bibfnamefont {P.}~\bibnamefont {Robichaud}}, \bibinfo {author} {\bibfnamefont {A.}~\bibnamefont {Winnicki}}, \bibinfo {author} {\bibfnamefont {C.}~\bibnamefont {Zhang}}, \bibinfo {author} {\bibfnamefont {L.}~\bibnamefont {Cheng}},\ and\ \bibinfo {author} {\bibfnamefont {J.~M.}\ \bibnamefont {Doyle}},\ }\href {https://link.aps.org/doi/10.1103/PhysRevLett.130.153202} {\bibfield  {journal} {\bibinfo  {journal} {Phys. Rev. Lett.}\ }\textbf {\bibinfo {volume} {130}},\ \bibinfo {pages} {153202} (\bibinfo {year} {2023})}\BibitemShut {NoStop}%
\bibitem [{\citenamefont {Kozyryev}\ \emph {et~al.}(2016)\citenamefont {Kozyryev}, \citenamefont {Baum}, \citenamefont {Matsuda},\ and\ \citenamefont {Doyle}}]{Kozyryev2016}%
  \BibitemOpen
  \bibfield  {author} {\bibinfo {author} {\bibfnamefont {I.}~\bibnamefont {Kozyryev}}, \bibinfo {author} {\bibfnamefont {L.}~\bibnamefont {Baum}}, \bibinfo {author} {\bibfnamefont {K.}~\bibnamefont {Matsuda}},\ and\ \bibinfo {author} {\bibfnamefont {J.~M.}\ \bibnamefont {Doyle}},\ }\href {https://chemistry-europe.onlinelibrary.wiley.com/doi/abs/10.1002/cphc.201601051} {\bibfield  {journal} {\bibinfo  {journal} {Chem. Phys. Chem.}\ }\textbf {\bibinfo {volume} {17}},\ \bibinfo {pages} {3641} (\bibinfo {year} {2016})}\BibitemShut {NoStop}%
\bibitem [{\citenamefont {Augenbraun}\ \emph {et~al.}(2020{\natexlab{a}})\citenamefont {Augenbraun}, \citenamefont {Doyle}, \citenamefont {Zelevinsky},\ and\ \citenamefont {Kozyryev}}]{Augenbraun2020}%
  \BibitemOpen
  \bibfield  {author} {\bibinfo {author} {\bibfnamefont {B.~L.}\ \bibnamefont {Augenbraun}}, \bibinfo {author} {\bibfnamefont {J.~M.}\ \bibnamefont {Doyle}}, \bibinfo {author} {\bibfnamefont {T.}~\bibnamefont {Zelevinsky}},\ and\ \bibinfo {author} {\bibfnamefont {I.}~\bibnamefont {Kozyryev}},\ }\href {https://journals.aps.org/prx/abstract/10.1103/PhysRevX.10.031022} {\bibfield  {journal} {\bibinfo  {journal} {Phys. Rev. X}\ }\textbf {\bibinfo {volume} {10}},\ \bibinfo {pages} {031022} (\bibinfo {year} {2020}{\natexlab{a}})}\BibitemShut {NoStop}%
\bibitem [{\citenamefont {Mitra}\ \emph {et~al.}(2020)\citenamefont {Mitra}, \citenamefont {Vilas}, \citenamefont {Hallas}, \citenamefont {Anderegg}, \citenamefont {Augenbraun}, \citenamefont {Baum}, \citenamefont {Miller}, \citenamefont {Raval},\ and\ \citenamefont {Doyle}}]{Mitra2020}%
  \BibitemOpen
  \bibfield  {author} {\bibinfo {author} {\bibfnamefont {D.}~\bibnamefont {Mitra}}, \bibinfo {author} {\bibfnamefont {N.~B.}\ \bibnamefont {Vilas}}, \bibinfo {author} {\bibfnamefont {C.}~\bibnamefont {Hallas}}, \bibinfo {author} {\bibfnamefont {L.}~\bibnamefont {Anderegg}}, \bibinfo {author} {\bibfnamefont {B.~L.}\ \bibnamefont {Augenbraun}}, \bibinfo {author} {\bibfnamefont {L.}~\bibnamefont {Baum}}, \bibinfo {author} {\bibfnamefont {C.}~\bibnamefont {Miller}}, \bibinfo {author} {\bibfnamefont {S.}~\bibnamefont {Raval}},\ and\ \bibinfo {author} {\bibfnamefont {J.~M.}\ \bibnamefont {Doyle}},\ }\href {https://www.doi.org/10.1126/science.abc5357} {\bibfield  {journal} {\bibinfo  {journal} {Science}\ }\textbf {\bibinfo {volume} {369}},\ \bibinfo {pages} {1366} (\bibinfo {year} {2020})}\BibitemShut {NoStop}%
\bibitem [{\citenamefont {Burchesky}(2023)}]{Burchesky2023}%
  \BibitemOpen
  \bibfield  {author} {\bibinfo {author} {\bibfnamefont {S.}~\bibnamefont {Burchesky}},\ }\emph {\bibinfo {title} {{Engineered Collisions, Molecular Qubits, and Laser Cooling of Asymmetric Top Molecules}}},\ \href {https://dash.harvard.edu/handle/1/37375823} {Ph.D. thesis},\ \bibinfo  {school} {Harvard University} (\bibinfo {year} {2023})\BibitemShut {NoStop}%
\bibitem [{\citenamefont {Flambaum}\ and\ \citenamefont {Khriplovich}(1985)}]{Flambaum1985}%
  \BibitemOpen
  \bibfield  {author} {\bibinfo {author} {\bibfnamefont {V.~V.}\ \bibnamefont {Flambaum}}\ and\ \bibinfo {author} {\bibfnamefont {I.~B.}\ \bibnamefont {Khriplovich}},\ }\href {https://www.sciencedirect.com/science/article/abs/pii/037596018590756X} {\bibfield  {journal} {\bibinfo  {journal} {Phys. Lett. A}\ }\textbf {\bibinfo {volume} {110}},\ \bibinfo {pages} {121} (\bibinfo {year} {1985})}\BibitemShut {NoStop}%
\bibitem [{\citenamefont {Dzuba}\ \emph {et~al.}(2017)\citenamefont {Dzuba}, \citenamefont {Flambaum},\ and\ \citenamefont {Stadnik}}]{Dzuba2017}%
  \BibitemOpen
  \bibfield  {author} {\bibinfo {author} {\bibfnamefont {V.~A.}\ \bibnamefont {Dzuba}}, \bibinfo {author} {\bibfnamefont {V.~V.}\ \bibnamefont {Flambaum}},\ and\ \bibinfo {author} {\bibfnamefont {Y.~V.}\ \bibnamefont {Stadnik}},\ }\href {https://link.aps.org/doi/10.1103/PhysRevLett.119.223201} {\bibfield  {journal} {\bibinfo  {journal} {Phys. Rev. Lett.}\ }\textbf {\bibinfo {volume} {119}},\ \bibinfo {pages} {223201} (\bibinfo {year} {2017})}\BibitemShut {NoStop}%
\bibitem [{\citenamefont {Wei}\ \emph {et~al.}(2011)\citenamefont {Wei}, \citenamefont {Kais}, \citenamefont {Friedrich},\ and\ \citenamefont {Herschbach}}]{Wei2011}%
  \BibitemOpen
  \bibfield  {author} {\bibinfo {author} {\bibfnamefont {Q.}~\bibnamefont {Wei}}, \bibinfo {author} {\bibfnamefont {S.}~\bibnamefont {Kais}}, \bibinfo {author} {\bibfnamefont {B.}~\bibnamefont {Friedrich}},\ and\ \bibinfo {author} {\bibfnamefont {D.}~\bibnamefont {Herschbach}},\ }\href {https://pubs.aip.org/aip/jcp/article/135/15/154102/190300/Entanglement-of-polar-symmetric-top-molecules-as} {\bibfield  {journal} {\bibinfo  {journal} {J. Chem. Phys.}\ }\textbf {\bibinfo {volume} {135}},\ \bibinfo {pages} {154102} (\bibinfo {year} {2011})}\BibitemShut {NoStop}%
\bibitem [{\citenamefont {Wall}\ \emph {et~al.}(2013)\citenamefont {Wall}, \citenamefont {Maeda},\ and\ \citenamefont {Carr}}]{Wall2013}%
  \BibitemOpen
  \bibfield  {author} {\bibinfo {author} {\bibfnamefont {M.~L.}\ \bibnamefont {Wall}}, \bibinfo {author} {\bibfnamefont {K.}~\bibnamefont {Maeda}},\ and\ \bibinfo {author} {\bibfnamefont {L.~D.}\ \bibnamefont {Carr}},\ }\href {https://doi.org/10.1002/andp.201300105} {\bibfield  {journal} {\bibinfo  {journal} {Ann. Phys. (Leipzig)}\ }\textbf {\bibinfo {volume} {525}},\ \bibinfo {pages} {845} (\bibinfo {year} {2013})}\BibitemShut {NoStop}%
\bibitem [{\citenamefont {Wall}\ \emph {et~al.}(2015)\citenamefont {Wall}, \citenamefont {Maeda},\ and\ \citenamefont {Carr}}]{Wall2015}%
  \BibitemOpen
  \bibfield  {author} {\bibinfo {author} {\bibfnamefont {M.~L.}\ \bibnamefont {Wall}}, \bibinfo {author} {\bibfnamefont {K.}~\bibnamefont {Maeda}},\ and\ \bibinfo {author} {\bibfnamefont {L.~D.}\ \bibnamefont {Carr}},\ }\href {https://iopscience.iop.org/article/10.1088/1367-2630/17/2/025001/meta} {\bibfield  {journal} {\bibinfo  {journal} {New J. Phys.}\ }\textbf {\bibinfo {volume} {17}},\ \bibinfo {pages} {025001} (\bibinfo {year} {2015})}\BibitemShut {NoStop}%
\bibitem [{\citenamefont {Yu}\ \emph {et~al.}(2019)\citenamefont {Yu}, \citenamefont {Cheuk}, \citenamefont {Kozyryev},\ and\ \citenamefont {Doyle}}]{Yu2019}%
  \BibitemOpen
  \bibfield  {author} {\bibinfo {author} {\bibfnamefont {P.}~\bibnamefont {Yu}}, \bibinfo {author} {\bibfnamefont {L.~W.}\ \bibnamefont {Cheuk}}, \bibinfo {author} {\bibfnamefont {I.}~\bibnamefont {Kozyryev}},\ and\ \bibinfo {author} {\bibfnamefont {J.~M.}\ \bibnamefont {Doyle}},\ }\href {https://iopscience.iop.org/article/10.1088/1367-2630/ab428d/meta} {\bibfield  {journal} {\bibinfo  {journal} {New J. Phys.}\ }\textbf {\bibinfo {volume} {21}},\ \bibinfo {pages} {093049} (\bibinfo {year} {2019})}\BibitemShut {NoStop}%
\bibitem [{\citenamefont {Balakrishnan}(2016)}]{Balakrishnan2016}%
  \BibitemOpen
  \bibfield  {author} {\bibinfo {author} {\bibfnamefont {N.}~\bibnamefont {Balakrishnan}},\ }\href {https://pubs.aip.org/aip/jcp/article/145/15/150901/935594/Perspective-Ultracold-molecules-and-the-dawn-of} {\bibfield  {journal} {\bibinfo  {journal} {J. Chem. Phys.}\ }\textbf {\bibinfo {volume} {145}},\ \bibinfo {pages} {150901} (\bibinfo {year} {2016})}\BibitemShut {NoStop}%
\bibitem [{\citenamefont {Bohn}\ \emph {et~al.}(2017)\citenamefont {Bohn}, \citenamefont {Rey},\ and\ \citenamefont {Ye}}]{Bohn2017}%
  \BibitemOpen
  \bibfield  {author} {\bibinfo {author} {\bibfnamefont {J.~L.}\ \bibnamefont {Bohn}}, \bibinfo {author} {\bibfnamefont {A.~M.}\ \bibnamefont {Rey}},\ and\ \bibinfo {author} {\bibfnamefont {J.}~\bibnamefont {Ye}},\ }\href {https://www.science.org/doi/10.1126/science.aam6299} {\bibfield  {journal} {\bibinfo  {journal} {Science}\ }\textbf {\bibinfo {volume} {357}},\ \bibinfo {pages} {1002} (\bibinfo {year} {2017})}\BibitemShut {NoStop}%
\bibitem [{\citenamefont {Li}\ \emph {et~al.}(2019)\citenamefont {Li}, \citenamefont {K{\l}os}, \citenamefont {Petrov},\ and\ \citenamefont {Kotochigova}}]{Li2019}%
  \BibitemOpen
  \bibfield  {author} {\bibinfo {author} {\bibfnamefont {M.}~\bibnamefont {Li}}, \bibinfo {author} {\bibfnamefont {J.}~\bibnamefont {K{\l}os}}, \bibinfo {author} {\bibfnamefont {A.}~\bibnamefont {Petrov}},\ and\ \bibinfo {author} {\bibfnamefont {S.}~\bibnamefont {Kotochigova}},\ }\href {https://www.nature.com/articles/s42005-019-0245-2} {\bibfield  {journal} {\bibinfo  {journal} {Commun. Phys.}\ }\textbf {\bibinfo {volume} {2}},\ \bibinfo {pages} {148} (\bibinfo {year} {2019})}\BibitemShut {NoStop}%
\bibitem [{\citenamefont {Paul}\ \emph {et~al.}(2019)\citenamefont {Paul}, \citenamefont {Sharma}, \citenamefont {Reza}, \citenamefont {Telfah}, \citenamefont {Miller},\ and\ \citenamefont {Liu}}]{Paul2019}%
  \BibitemOpen
  \bibfield  {author} {\bibinfo {author} {\bibfnamefont {A.~C.}\ \bibnamefont {Paul}}, \bibinfo {author} {\bibfnamefont {K.}~\bibnamefont {Sharma}}, \bibinfo {author} {\bibfnamefont {M.~A.}\ \bibnamefont {Reza}}, \bibinfo {author} {\bibfnamefont {H.}~\bibnamefont {Telfah}}, \bibinfo {author} {\bibfnamefont {T.~A.}\ \bibnamefont {Miller}},\ and\ \bibinfo {author} {\bibfnamefont {J.}~\bibnamefont {Liu}},\ }\href {https://pubs.aip.org/aip/jcp/article/151/13/134303/313421/Laser-induced-fluorescence-and-dispersed} {\bibfield  {journal} {\bibinfo  {journal} {J. Chem. Phys.}\ }\textbf {\bibinfo {volume} {151}},\ \bibinfo {pages} {134303} (\bibinfo {year} {2019})}\BibitemShut {NoStop}%
\bibitem [{\citenamefont {Mengesha}\ \emph {et~al.}(2020)\citenamefont {Mengesha}, \citenamefont {Le}, \citenamefont {Steimle}, \citenamefont {Cheng}, \citenamefont {Zhang}, \citenamefont {Augenbraun}, \citenamefont {Lasner},\ and\ \citenamefont {Doyle}}]{Mengesha2020}%
  \BibitemOpen
  \bibfield  {author} {\bibinfo {author} {\bibfnamefont {E.~T.}\ \bibnamefont {Mengesha}}, \bibinfo {author} {\bibfnamefont {A.~T.}\ \bibnamefont {Le}}, \bibinfo {author} {\bibfnamefont {T.~C.}\ \bibnamefont {Steimle}}, \bibinfo {author} {\bibfnamefont {L.}~\bibnamefont {Cheng}}, \bibinfo {author} {\bibfnamefont {C.}~\bibnamefont {Zhang}}, \bibinfo {author} {\bibfnamefont {B.~L.}\ \bibnamefont {Augenbraun}}, \bibinfo {author} {\bibfnamefont {Z.}~\bibnamefont {Lasner}},\ and\ \bibinfo {author} {\bibfnamefont {J.}~\bibnamefont {Doyle}},\ }\href {https://pubs.acs.org/doi/full/10.1021/acs.jpca.0c00850} {\bibfield  {journal} {\bibinfo  {journal} {J. Phys. Chem. A}\ }\textbf {\bibinfo {volume} {124}},\ \bibinfo {pages} {3135} (\bibinfo {year} {2020})}\BibitemShut {NoStop}%
\bibitem [{\citenamefont {K{\l}os}\ and\ \citenamefont {Kotochigova}(2020)}]{Kos20}%
  \BibitemOpen
  \bibfield  {author} {\bibinfo {author} {\bibfnamefont {J.}~\bibnamefont {K{\l}os}}\ and\ \bibinfo {author} {\bibfnamefont {S.}~\bibnamefont {Kotochigova}},\ }\href {https://link.aps.org/doi/10.1103/PhysRevResearch.2.013384} {\bibfield  {journal} {\bibinfo  {journal} {Phys. Rev. Res.}\ }\textbf {\bibinfo {volume} {2}},\ \bibinfo {pages} {13384} (\bibinfo {year} {2020})}\BibitemShut {NoStop}%
\bibitem [{\citenamefont {Zhang}\ \emph {et~al.}(2021)\citenamefont {Zhang}, \citenamefont {Augenbraun}, \citenamefont {Lasner}, \citenamefont {Vilas}, \citenamefont {Doyle},\ and\ \citenamefont {Cheng}}]{Zhang2021}%
  \BibitemOpen
  \bibfield  {author} {\bibinfo {author} {\bibfnamefont {C.}~\bibnamefont {Zhang}}, \bibinfo {author} {\bibfnamefont {B.}~\bibnamefont {Augenbraun}}, \bibinfo {author} {\bibfnamefont {Z.~D.}\ \bibnamefont {Lasner}}, \bibinfo {author} {\bibfnamefont {N.~B.}\ \bibnamefont {Vilas}}, \bibinfo {author} {\bibfnamefont {J.~M.}\ \bibnamefont {Doyle}},\ and\ \bibinfo {author} {\bibfnamefont {L.}~\bibnamefont {Cheng}},\ }\href {https://pubs.aip.org/aip/jcp/article/155/9/091101/199807/Accurate-prediction-and-measurement-of-vibronic} {\bibfield  {journal} {\bibinfo  {journal} {J. Chem. Phys.}\ }\textbf {\bibinfo {volume} {155}},\ \bibinfo {pages} {91101} (\bibinfo {year} {2021})}\BibitemShut {NoStop}%
\bibitem [{\citenamefont {Lasner}\ \emph {et~al.}(2022)\citenamefont {Lasner}, \citenamefont {Lunstad}, \citenamefont {Zhang}, \citenamefont {Cheng},\ and\ \citenamefont {Doyle}}]{Lasner2022}%
  \BibitemOpen
  \bibfield  {author} {\bibinfo {author} {\bibfnamefont {Z.}~\bibnamefont {Lasner}}, \bibinfo {author} {\bibfnamefont {A.}~\bibnamefont {Lunstad}}, \bibinfo {author} {\bibfnamefont {C.}~\bibnamefont {Zhang}}, \bibinfo {author} {\bibfnamefont {L.}~\bibnamefont {Cheng}},\ and\ \bibinfo {author} {\bibfnamefont {J.~M.}\ \bibnamefont {Doyle}},\ }\href {https://journals.aps.org/pra/abstract/10.1103/PhysRevA.106.L020801} {\bibfield  {journal} {\bibinfo  {journal} {Phys. Rev. A}\ }\textbf {\bibinfo {volume} {106}},\ \bibinfo {pages} {20801} (\bibinfo {year} {2022})}\BibitemShut {NoStop}%
\bibitem [{\citenamefont {Zhang}\ \emph {et~al.}(2023{\natexlab{a}})\citenamefont {Zhang}, \citenamefont {Hutzler},\ and\ \citenamefont {Cheng}}]{Zhang23}%
  \BibitemOpen
  \bibfield  {author} {\bibinfo {author} {\bibfnamefont {C.}~\bibnamefont {Zhang}}, \bibinfo {author} {\bibfnamefont {N.~R.}\ \bibnamefont {Hutzler}},\ and\ \bibinfo {author} {\bibfnamefont {L.}~\bibnamefont {Cheng}},\ }\href {https://doi.org/10.1021/acs.jctc.3c00408} {\bibfield  {journal} {\bibinfo  {journal} {J. Chem. Theory Comput.}\ }\textbf {\bibinfo {volume} {19}},\ \bibinfo {pages} {4136} (\bibinfo {year} {2023}{\natexlab{a}})}\BibitemShut {NoStop}%
\bibitem [{\citenamefont {Zhu}\ \emph {et~al.}(2024)\citenamefont {Zhu}, \citenamefont {Lao}, \citenamefont {Dickerson}, \citenamefont {Caram}, \citenamefont {Campbell}, \citenamefont {Alexandrova},\ and\ \citenamefont {Hudson}}]{Zhu24}%
  \BibitemOpen
  \bibfield  {author} {\bibinfo {author} {\bibfnamefont {G.-Z.}\ \bibnamefont {Zhu}}, \bibinfo {author} {\bibfnamefont {G.}~\bibnamefont {Lao}}, \bibinfo {author} {\bibfnamefont {C.~E.}\ \bibnamefont {Dickerson}}, \bibinfo {author} {\bibfnamefont {J.~R.}\ \bibnamefont {Caram}}, \bibinfo {author} {\bibfnamefont {W.~C.}\ \bibnamefont {Campbell}}, \bibinfo {author} {\bibfnamefont {A.~N.}\ \bibnamefont {Alexandrova}},\ and\ \bibinfo {author} {\bibfnamefont {E.~R.}\ \bibnamefont {Hudson}},\ }\href {https://doi.org/10.1021/acs.jpclett.3c03177} {\bibfield  {journal} {\bibinfo  {journal} {J. Phys. Chem. Lett.}\ }\textbf {\bibinfo {volume} {15}},\ \bibinfo {pages} {590} (\bibinfo {year} {2024})}\BibitemShut {NoStop}%
\bibitem [{\citenamefont {Bethlem}\ and\ \citenamefont {Meijer}(2003)}]{Bethlem2003}%
  \BibitemOpen
  \bibfield  {author} {\bibinfo {author} {\bibfnamefont {H.~L.}\ \bibnamefont {Bethlem}}\ and\ \bibinfo {author} {\bibfnamefont {G.}~\bibnamefont {Meijer}},\ }\href {https://www.tandfonline.com/doi/abs/10.1080/0144235021000046422} {\bibfield  {journal} {\bibinfo  {journal} {Int. Rev. Phys. Chem.}\ }\textbf {\bibinfo {volume} {22}},\ \bibinfo {pages} {73} (\bibinfo {year} {2003})}\BibitemShut {NoStop}%
\bibitem [{\citenamefont {Stuhl}\ \emph {et~al.}(2008)\citenamefont {Stuhl}, \citenamefont {Sawyer}, \citenamefont {Wang},\ and\ \citenamefont {Ye}}]{Stuhl2008}%
  \BibitemOpen
  \bibfield  {author} {\bibinfo {author} {\bibfnamefont {B.~K.}\ \bibnamefont {Stuhl}}, \bibinfo {author} {\bibfnamefont {B.~C.}\ \bibnamefont {Sawyer}}, \bibinfo {author} {\bibfnamefont {D.}~\bibnamefont {Wang}},\ and\ \bibinfo {author} {\bibfnamefont {J.}~\bibnamefont {Ye}},\ }\href {https://journals.aps.org/prl/abstract/10.1103/PhysRevLett.101.243002} {\bibfield  {journal} {\bibinfo  {journal} {Phys. Rev. Lett.}\ }\textbf {\bibinfo {volume} {101}},\ \bibinfo {pages} {243002} (\bibinfo {year} {2008})}\BibitemShut {NoStop}%
\bibitem [{\citenamefont {{Di Rosa}}(2004)}]{DiRosa2004}%
  \BibitemOpen
  \bibfield  {author} {\bibinfo {author} {\bibfnamefont {M.~D.}\ \bibnamefont {{Di Rosa}}},\ }\href {https://link.springer.com/article/10.1140/epjd/e2004-00167-2} {\bibfield  {journal} {\bibinfo  {journal} {Eur. Phys. J. D}\ }\textbf {\bibinfo {volume} {31}},\ \bibinfo {pages} {395} (\bibinfo {year} {2004})}\BibitemShut {NoStop}%
\bibitem [{\citenamefont {Kozyryev}\ \emph {et~al.}(2019)\citenamefont {Kozyryev}, \citenamefont {Steimle}, \citenamefont {Yu}, \citenamefont {Nguyen},\ and\ \citenamefont {Doyle}}]{Kozyryev2019}%
  \BibitemOpen
  \bibfield  {author} {\bibinfo {author} {\bibfnamefont {I.}~\bibnamefont {Kozyryev}}, \bibinfo {author} {\bibfnamefont {T.~C.}\ \bibnamefont {Steimle}}, \bibinfo {author} {\bibfnamefont {P.}~\bibnamefont {Yu}}, \bibinfo {author} {\bibfnamefont {D.-T.}\ \bibnamefont {Nguyen}},\ and\ \bibinfo {author} {\bibfnamefont {J.~M.}\ \bibnamefont {Doyle}},\ }\href {https://doi.org/10.1088/1367-2630/ab19d7} {\bibfield  {journal} {\bibinfo  {journal} {New J. Phys}\ }\textbf {\bibinfo {volume} {21}},\ \bibinfo {pages} {52002} (\bibinfo {year} {2019})}\BibitemShut {NoStop}%
\bibitem [{\citenamefont {Stanton}\ and\ \citenamefont {Bartlett}(1993)}]{Stanton93a}%
  \BibitemOpen
  \bibfield  {author} {\bibinfo {author} {\bibfnamefont {J.~F.}\ \bibnamefont {Stanton}}\ and\ \bibinfo {author} {\bibfnamefont {R.~J.}\ \bibnamefont {Bartlett}},\ }\href {https://pubs.aip.org/aip/jcp/article-abstract/98/9/7029/857758/The-equation-of-motion-coupled-cluster-method-A?redirectedFrom=fulltext} {\bibfield  {journal} {\bibinfo  {journal} {J. Chem. Phys.}\ }\textbf {\bibinfo {volume} {98}},\ \bibinfo {pages} {7029} (\bibinfo {year} {1993})}\BibitemShut {NoStop}%
\bibitem [{\citenamefont {Nooijen}\ and\ \citenamefont {Bartlett}(1995)}]{Nooijen95}%
  \BibitemOpen
  \bibfield  {author} {\bibinfo {author} {\bibfnamefont {M.}~\bibnamefont {Nooijen}}\ and\ \bibinfo {author} {\bibfnamefont {R.~J.}\ \bibnamefont {Bartlett}},\ }\href {https://pubs.aip.org/aip/jcp/article-abstract/102/9/3629/481399/Equation-of-motion-coupled-cluster-method-for?redirectedFrom=fulltext} {\bibfield  {journal} {\bibinfo  {journal} {J. Chem. Phys.}\ }\textbf {\bibinfo {volume} {102}},\ \bibinfo {pages} {3629} (\bibinfo {year} {1995})}\BibitemShut {NoStop}%
\bibitem [{\citenamefont {Rabidoux}\ \emph {et~al.}(2016)\citenamefont {Rabidoux}, \citenamefont {Eijkhout},\ and\ \citenamefont {Stanton}}]{Rabidoux16}%
  \BibitemOpen
  \bibfield  {author} {\bibinfo {author} {\bibfnamefont {S.~M.}\ \bibnamefont {Rabidoux}}, \bibinfo {author} {\bibfnamefont {V.}~\bibnamefont {Eijkhout}},\ and\ \bibinfo {author} {\bibfnamefont {J.~F.}\ \bibnamefont {Stanton}},\ }\href {https://doi.org/10.1021/acs.jctc.5b00560} {\bibfield  {journal} {\bibinfo  {journal} {J. Chem. Theory Comput.}\ }\textbf {\bibinfo {volume} {12}},\ \bibinfo {pages} {728} (\bibinfo {year} {2016})}\BibitemShut {NoStop}%
\bibitem [{\citenamefont {Stanton}\ \emph {et~al.}()\citenamefont {Stanton}, \citenamefont {Gauss}, \citenamefont {Cheng}, \citenamefont {Harding}, \citenamefont {Matthews},\ and\ \citenamefont {Szalay}}]{CFOURfull}%
  \BibitemOpen
  \bibfield  {author} {\bibinfo {author} {\bibfnamefont {J.~F.}\ \bibnamefont {Stanton}}, \bibinfo {author} {\bibfnamefont {J.}~\bibnamefont {Gauss}}, \bibinfo {author} {\bibfnamefont {L.}~\bibnamefont {Cheng}}, \bibinfo {author} {\bibfnamefont {M.~E.}\ \bibnamefont {Harding}}, \bibinfo {author} {\bibfnamefont {D.~A.}\ \bibnamefont {Matthews}},\ and\ \bibinfo {author} {\bibfnamefont {P.~G.}\ \bibnamefont {Szalay}},\ }\href@noop {} {\bibinfo {title} {{CFOUR, Coupled-Cluster techniques for Computational Chemistry, a quantum-chemical program package}}}\BibitemShut {NoStop}%
\bibitem [{\citenamefont {Matthews}\ \emph {et~al.}(2020)\citenamefont {Matthews}, \citenamefont {Cheng}, \citenamefont {Harding}, \citenamefont {Lipparini}, \citenamefont {Stopkowicz}, \citenamefont {Jagau}, \citenamefont {Szalay}, \citenamefont {Gauss},\ and\ \citenamefont {Stanton}}]{Matthews20a}%
  \BibitemOpen
  \bibfield  {author} {\bibinfo {author} {\bibfnamefont {D.~A.}\ \bibnamefont {Matthews}}, \bibinfo {author} {\bibfnamefont {L.}~\bibnamefont {Cheng}}, \bibinfo {author} {\bibfnamefont {M.~E.}\ \bibnamefont {Harding}}, \bibinfo {author} {\bibfnamefont {F.}~\bibnamefont {Lipparini}}, \bibinfo {author} {\bibfnamefont {S.}~\bibnamefont {Stopkowicz}}, \bibinfo {author} {\bibfnamefont {T.-C.}\ \bibnamefont {Jagau}}, \bibinfo {author} {\bibfnamefont {P.~G.}\ \bibnamefont {Szalay}}, \bibinfo {author} {\bibfnamefont {J.}~\bibnamefont {Gauss}},\ and\ \bibinfo {author} {\bibfnamefont {J.~F.}\ \bibnamefont {Stanton}},\ }\href {https://doi.org/10.1063/5.0004837} {\bibfield  {journal} {\bibinfo  {journal} {J. Chem. Phys.}\ }\textbf {\bibinfo {volume} {152}},\ \bibinfo {pages} {214108} (\bibinfo {year} {2020})}\BibitemShut {NoStop}%
\bibitem [{\citenamefont {Mills}(1972)}]{Mills72}%
  \BibitemOpen
  \bibfield  {author} {\bibinfo {author} {\bibfnamefont {I.~M.}\ \bibnamefont {Mills}},\ }in\ \href@noop {} {\emph {\bibinfo {booktitle} {Molecular Spectroscopy: Modern Research}}},\ \bibinfo {editor} {edited by\ \bibinfo {editor} {\bibfnamefont {K.~N.}\ \bibnamefont {Rao}}\ and\ \bibinfo {editor} {\bibfnamefont {C.~W.}\ \bibnamefont {Mathews}}}\ (\bibinfo  {publisher} {Academic Press},\ \bibinfo {address} {New York},\ \bibinfo {year} {1972})\ pp.\ \bibinfo {pages} {115--140}\BibitemShut {NoStop}%
\bibitem [{\citenamefont {Kjaergaard}\ \emph {et~al.}(2008)\citenamefont {Kjaergaard}, \citenamefont {Garden}, \citenamefont {Chaban}, \citenamefont {Gerber}, \citenamefont {Matthews},\ and\ \citenamefont {Stanton}}]{Kjaergaard08}%
  \BibitemOpen
  \bibfield  {author} {\bibinfo {author} {\bibfnamefont {H.~G.}\ \bibnamefont {Kjaergaard}}, \bibinfo {author} {\bibfnamefont {A.~L.}\ \bibnamefont {Garden}}, \bibinfo {author} {\bibfnamefont {G.~M.}\ \bibnamefont {Chaban}}, \bibinfo {author} {\bibfnamefont {R.~B.}\ \bibnamefont {Gerber}}, \bibinfo {author} {\bibfnamefont {D.~A.}\ \bibnamefont {Matthews}},\ and\ \bibinfo {author} {\bibfnamefont {J.~F.}\ \bibnamefont {Stanton}},\ }\href {https://doi.org/10.1021/jp710066f} {\bibfield  {journal} {\bibinfo  {journal} {J. Phys. Chem. A}\ }\textbf {\bibinfo {volume} {112}},\ \bibinfo {pages} {4324} (\bibinfo {year} {2008})}\BibitemShut {NoStop}%
\bibitem [{\citenamefont {Dyall}(1997)}]{Dyall97}%
  \BibitemOpen
  \bibfield  {author} {\bibinfo {author} {\bibfnamefont {K.~G.}\ \bibnamefont {Dyall}},\ }\href {https://doi.org/10.1063/1.473860} {\bibfield  {journal} {\bibinfo  {journal} {J. Chem. Phys.}\ }\textbf {\bibinfo {volume} {106}},\ \bibinfo {pages} {9618} (\bibinfo {year} {1997})}\BibitemShut {NoStop}%
\bibitem [{\citenamefont {Liu}\ and\ \citenamefont {Peng}(2009)}]{Liu09}%
  \BibitemOpen
  \bibfield  {author} {\bibinfo {author} {\bibfnamefont {W.}~\bibnamefont {Liu}}\ and\ \bibinfo {author} {\bibfnamefont {D.}~\bibnamefont {Peng}},\ }\href {https://pubs.aip.org/aip/jcp/article/131/3/031104/71694} {\bibfield  {journal} {\bibinfo  {journal} {J. Chem. Phys.}\ }\textbf {\bibinfo {volume} {131}},\ \bibinfo {pages} {031104} (\bibinfo {year} {2009})}\BibitemShut {NoStop}%
\bibitem [{\citenamefont {Dyall}(2001)}]{Dyall01}%
  \BibitemOpen
  \bibfield  {author} {\bibinfo {author} {\bibfnamefont {K.~G.}\ \bibnamefont {Dyall}},\ }\href {https://doi.org/10.1063/1.1413512} {\bibfield  {journal} {\bibinfo  {journal} {J. Chem. Phys.}\ }\textbf {\bibinfo {volume} {115}},\ \bibinfo {pages} {9136} (\bibinfo {year} {2001})}\BibitemShut {NoStop}%
\bibitem [{\citenamefont {Zhang}\ \emph {et~al.}(2023{\natexlab{b}})\citenamefont {Zhang}, \citenamefont {Zheng}, \citenamefont {Liu}, \citenamefont {Asthana},\ and\ \citenamefont {Cheng}}]{Zhang23a}%
  \BibitemOpen
  \bibfield  {author} {\bibinfo {author} {\bibfnamefont {C.}~\bibnamefont {Zhang}}, \bibinfo {author} {\bibfnamefont {X.}~\bibnamefont {Zheng}}, \bibinfo {author} {\bibfnamefont {J.}~\bibnamefont {Liu}}, \bibinfo {author} {\bibfnamefont {A.}~\bibnamefont {Asthana}},\ and\ \bibinfo {author} {\bibfnamefont {L.}~\bibnamefont {Cheng}},\ }\href {https://doi.org/10.1063/5.0175041} {\bibfield  {journal} {\bibinfo  {journal} {J. Chem. Phys.}\ }\textbf {\bibinfo {volume} {159}},\ \bibinfo {pages} {244113} (\bibinfo {year} {2023}{\natexlab{b}})}\BibitemShut {NoStop}%
\bibitem [{\citenamefont {Liu}\ and\ \citenamefont {Cheng}(2018)}]{Liu18}%
  \BibitemOpen
  \bibfield  {author} {\bibinfo {author} {\bibfnamefont {J.}~\bibnamefont {Liu}}\ and\ \bibinfo {author} {\bibfnamefont {L.}~\bibnamefont {Cheng}},\ }\href {https://doi.org/10.1063/1.5023750} {\bibfield  {journal} {\bibinfo  {journal} {J. Chem. Phys.}\ }\textbf {\bibinfo {volume} {148}},\ \bibinfo {pages} {144108} (\bibinfo {year} {2018})}\BibitemShut {NoStop}%
\bibitem [{\citenamefont {Zhang}\ and\ \citenamefont {Cheng}(2022)}]{Zhang22}%
  \BibitemOpen
  \bibfield  {author} {\bibinfo {author} {\bibfnamefont {C.}~\bibnamefont {Zhang}}\ and\ \bibinfo {author} {\bibfnamefont {L.}~\bibnamefont {Cheng}},\ }\href {https://doi.org/10.1021/acs.jpca.2c02181} {\bibfield  {journal} {\bibinfo  {journal} {J. Phys. Chem. A}\ }\textbf {\bibinfo {volume} {126}},\ \bibinfo {pages} {4537} (\bibinfo {year} {2022})}\BibitemShut {NoStop}%
\bibitem [{\citenamefont {Cheng}\ and\ \citenamefont {Gauss}(2011)}]{Cheng11b}%
  \BibitemOpen
  \bibfield  {author} {\bibinfo {author} {\bibfnamefont {L.}~\bibnamefont {Cheng}}\ and\ \bibinfo {author} {\bibfnamefont {J.}~\bibnamefont {Gauss}},\ }\href {https://doi.org/10.1063/1.3624397} {\bibfield  {journal} {\bibinfo  {journal} {J. Chem. Phys.}\ }\textbf {\bibinfo {volume} {135}},\ \bibinfo {pages} {084114} (\bibinfo {year} {2011})}\BibitemShut {NoStop}%
\bibitem [{\citenamefont {Hill}\ and\ \citenamefont {Peterson}(2017)}]{Hill17}%
  \BibitemOpen
  \bibfield  {author} {\bibinfo {author} {\bibfnamefont {J.~G.}\ \bibnamefont {Hill}}\ and\ \bibinfo {author} {\bibfnamefont {K.~A.}\ \bibnamefont {Peterson}},\ }\href {https://doi.org/10.1063/1.5010587} {\bibfield  {journal} {\bibinfo  {journal} {J. Chem. Phys.}\ }\textbf {\bibinfo {volume} {147}},\ \bibinfo {pages} {244106} (\bibinfo {year} {2017})}\BibitemShut {NoStop}%
\bibitem [{\citenamefont {Woon}\ and\ \citenamefont {{Dunning Jr.}}(1993)}]{Woon93}%
  \BibitemOpen
  \bibfield  {author} {\bibinfo {author} {\bibfnamefont {D.~E.}\ \bibnamefont {Woon}}\ and\ \bibinfo {author} {\bibfnamefont {T.~H.}\ \bibnamefont {{Dunning Jr.}}},\ }\href {https://doi.org/10.1063/1.464303} {\bibfield  {journal} {\bibinfo  {journal} {J. Chem. Phys.}\ }\textbf {\bibinfo {volume} {98}},\ \bibinfo {pages} {1358} (\bibinfo {year} {1993})}\BibitemShut {NoStop}%
\bibitem [{\citenamefont {Peterson}\ and\ \citenamefont {{Dunning Jr.}}(2002)}]{Peterson02}%
  \BibitemOpen
  \bibfield  {author} {\bibinfo {author} {\bibfnamefont {K.~A.}\ \bibnamefont {Peterson}}\ and\ \bibinfo {author} {\bibfnamefont {T.~H.}\ \bibnamefont {{Dunning Jr.}}},\ }\href {https://doi.org/10.1063/1.1520138} {\bibfield  {journal} {\bibinfo  {journal} {J. Chem. Phys.}\ }\textbf {\bibinfo {volume} {117}},\ \bibinfo {pages} {10548} (\bibinfo {year} {2002})}\BibitemShut {NoStop}%
\bibitem [{\citenamefont {{Dunning, Jr.}}(1989)}]{Dunning89}%
  \BibitemOpen
  \bibfield  {author} {\bibinfo {author} {\bibfnamefont {T.~H.}\ \bibnamefont {{Dunning, Jr.}}},\ }\href {https://doi.org/10.1063/1.456153} {\bibfield  {journal} {\bibinfo  {journal} {J. Chem. Phys.}\ }\textbf {\bibinfo {volume} {90}},\ \bibinfo {pages} {1007} (\bibinfo {year} {1989})}\BibitemShut {NoStop}%
\bibitem [{\citenamefont {F{\ae}gri}(2001)}]{Faegri01}%
  \BibitemOpen
  \bibfield  {author} {\bibinfo {author} {\bibfnamefont {K.}~\bibnamefont {F{\ae}gri}},\ }\href {https://doi.org/10.1007/s002140000209} {\bibfield  {journal} {\bibinfo  {journal} {Theor. Chem. Acc.}\ }\textbf {\bibinfo {volume} {105}},\ \bibinfo {pages} {252} (\bibinfo {year} {2001})}\BibitemShut {NoStop}%
\bibitem [{\citenamefont {Roos}\ \emph {et~al.}(2004{\natexlab{a}})\citenamefont {Roos}, \citenamefont {Veryazov},\ and\ \citenamefont {Widmark}}]{Roos04a}%
  \BibitemOpen
  \bibfield  {author} {\bibinfo {author} {\bibfnamefont {B.~O.}\ \bibnamefont {Roos}}, \bibinfo {author} {\bibfnamefont {V.}~\bibnamefont {Veryazov}},\ and\ \bibinfo {author} {\bibfnamefont {P.-O.}\ \bibnamefont {Widmark}},\ }\href {https://doi.org/10.1007/s00214-003-0537-0} {\bibfield  {journal} {\bibinfo  {journal} {Theor. Chem. Acc.}\ }\textbf {\bibinfo {volume} {111}},\ \bibinfo {pages} {345} (\bibinfo {year} {2004}{\natexlab{a}})}\BibitemShut {NoStop}%
\bibitem [{\citenamefont {Roos}\ \emph {et~al.}(2004{\natexlab{b}})\citenamefont {Roos}, \citenamefont {Lindh}, \citenamefont {Malmqvist}, \citenamefont {Veryazov},\ and\ \citenamefont {Widmark}}]{Roos04}%
  \BibitemOpen
  \bibfield  {author} {\bibinfo {author} {\bibfnamefont {B.~O.}\ \bibnamefont {Roos}}, \bibinfo {author} {\bibfnamefont {R.}~\bibnamefont {Lindh}}, \bibinfo {author} {\bibfnamefont {P.-{\AA}.}\ \bibnamefont {Malmqvist}}, \bibinfo {author} {\bibfnamefont {V.}~\bibnamefont {Veryazov}},\ and\ \bibinfo {author} {\bibfnamefont {P.-O.}\ \bibnamefont {Widmark}},\ }\href {https://doi.org/10.1021/jp031064+} {\bibfield  {journal} {\bibinfo  {journal} {J. Phys. Chem. A}\ }\textbf {\bibinfo {volume} {108}},\ \bibinfo {pages} {2851} (\bibinfo {year} {2004}{\natexlab{b}})}\BibitemShut {NoStop}%
\bibitem [{\citenamefont {Wormsbecher}\ and\ \citenamefont {Suenram}(1982)}]{Wormsbecher1982}%
  \BibitemOpen
  \bibfield  {author} {\bibinfo {author} {\bibfnamefont {R.~F.}\ \bibnamefont {Wormsbecher}}\ and\ \bibinfo {author} {\bibfnamefont {R.~D.}\ \bibnamefont {Suenram}},\ }\href {https://doi.org/10.1016/0022-2852(82)90138-2} {\bibfield  {journal} {\bibinfo  {journal} {J. Mol. Spect.}\ }\textbf {\bibinfo {volume} {95}},\ \bibinfo {pages} {391} (\bibinfo {year} {1982})}\BibitemShut {NoStop}%
\bibitem [{\citenamefont {O'Brien}\ \emph {et~al.}(1988)\citenamefont {O'Brien}, \citenamefont {Brazier},\ and\ \citenamefont {Bernath}}]{OBrien1988}%
  \BibitemOpen
  \bibfield  {author} {\bibinfo {author} {\bibfnamefont {L.~C.}\ \bibnamefont {O'Brien}}, \bibinfo {author} {\bibfnamefont {C.~R.}\ \bibnamefont {Brazier}},\ and\ \bibinfo {author} {\bibfnamefont {P.~F.}\ \bibnamefont {Bernath}},\ }\href {https://doi.org/10.1016/0022-2852(88)90280-9} {\bibfield  {journal} {\bibinfo  {journal} {J. Mol. Spect.}\ }\textbf {\bibinfo {volume} {130}},\ \bibinfo {pages} {13} (\bibinfo {year} {1988})}\BibitemShut {NoStop}%
\bibitem [{\citenamefont {Forthomme}\ \emph {et~al.}(2011)\citenamefont {Forthomme}, \citenamefont {Linton}, \citenamefont {Read}, \citenamefont {Tokaryk}, \citenamefont {Adam}, \citenamefont {Downie}, \citenamefont {Granger},\ and\ \citenamefont {Hopkins}}]{Forthomme2011}%
  \BibitemOpen
  \bibfield  {author} {\bibinfo {author} {\bibfnamefont {D.}~\bibnamefont {Forthomme}}, \bibinfo {author} {\bibfnamefont {C.}~\bibnamefont {Linton}}, \bibinfo {author} {\bibfnamefont {A.~G.}\ \bibnamefont {Read}}, \bibinfo {author} {\bibfnamefont {D.~W.}\ \bibnamefont {Tokaryk}}, \bibinfo {author} {\bibfnamefont {A.~G.}\ \bibnamefont {Adam}}, \bibinfo {author} {\bibfnamefont {L.~E.}\ \bibnamefont {Downie}}, \bibinfo {author} {\bibfnamefont {A.~D.}\ \bibnamefont {Granger}},\ and\ \bibinfo {author} {\bibfnamefont {W.~S.}\ \bibnamefont {Hopkins}},\ }\href {https://doi.org/10.1016/j.jms.2011.09.008} {\bibfield  {journal} {\bibinfo  {journal} {J. Mol. Spect.}\ }\textbf {\bibinfo {volume} {270}},\ \bibinfo {pages} {108} (\bibinfo {year} {2011})}\BibitemShut {NoStop}%
\bibitem [{\citenamefont {Butcher}\ \emph {et~al.}(1993)\citenamefont {Butcher}, \citenamefont {Chardonnet},\ and\ \citenamefont {Borde}}]{Butcher1993}%
  \BibitemOpen
  \bibfield  {author} {\bibinfo {author} {\bibfnamefont {R.}~\bibnamefont {Butcher}}, \bibinfo {author} {\bibfnamefont {C.}~\bibnamefont {Chardonnet}},\ and\ \bibinfo {author} {\bibfnamefont {C.}~\bibnamefont {Borde}},\ }\href {https://doi.org/10.1103/PhysRevLett.70.2698} {\bibfield  {journal} {\bibinfo  {journal} {Phys. Rev. Lett.}\ }\textbf {\bibinfo {volume} {70}},\ \bibinfo {pages} {2698} (\bibinfo {year} {1993})}\BibitemShut {NoStop}%
\bibitem [{\citenamefont {Brazier}\ and\ \citenamefont {Bernath}(1989)}]{Brazier1989}%
  \BibitemOpen
  \bibfield  {author} {\bibinfo {author} {\bibfnamefont {C.~R.}\ \bibnamefont {Brazier}}\ and\ \bibinfo {author} {\bibfnamefont {P.~F.}\ \bibnamefont {Bernath}},\ }\href {https://doi.org/10.1063/1.456742} {\bibfield  {journal} {\bibinfo  {journal} {J. Chem. Phys.}\ }\textbf {\bibinfo {volume} {91}},\ \bibinfo {pages} {4548} (\bibinfo {year} {1989})}\BibitemShut {NoStop}%
\bibitem [{\citenamefont {Marr}\ \emph {et~al.}(1996)\citenamefont {Marr}, \citenamefont {Grieman},\ and\ \citenamefont {Steimle}}]{Marr1996}%
  \BibitemOpen
  \bibfield  {author} {\bibinfo {author} {\bibfnamefont {A.~J.}\ \bibnamefont {Marr}}, \bibinfo {author} {\bibfnamefont {F.}~\bibnamefont {Grieman}},\ and\ \bibinfo {author} {\bibfnamefont {T.~C.}\ \bibnamefont {Steimle}},\ }\href {https://doi.org/10.1063/1.472265} {\bibfield  {journal} {\bibinfo  {journal} {J. Chem. Phys.}\ }\textbf {\bibinfo {volume} {105}},\ \bibinfo {pages} {3930} (\bibinfo {year} {1996})}\BibitemShut {NoStop}%
\bibitem [{\citenamefont {Wormsbecher}\ \emph {et~al.}(1983)\citenamefont {Wormsbecher}, \citenamefont {Trkula}, \citenamefont {Martner}, \citenamefont {Penn},\ and\ \citenamefont {Harris}}]{Wormsbecher1983}%
  \BibitemOpen
  \bibfield  {author} {\bibinfo {author} {\bibfnamefont {R.~F.}\ \bibnamefont {Wormsbecher}}, \bibinfo {author} {\bibfnamefont {M.}~\bibnamefont {Trkula}}, \bibinfo {author} {\bibfnamefont {C.}~\bibnamefont {Martner}}, \bibinfo {author} {\bibfnamefont {R.~E.}\ \bibnamefont {Penn}},\ and\ \bibinfo {author} {\bibfnamefont {D.~O.}\ \bibnamefont {Harris}},\ }\href {https://doi.org/10.1016/0022-2852(83)90335-1} {\bibfield  {journal} {\bibinfo  {journal} {J. Mol. Spect.}\ }\textbf {\bibinfo {volume} {97}},\ \bibinfo {pages} {29} (\bibinfo {year} {1983})}\BibitemShut {NoStop}%
\bibitem [{\citenamefont {Bopegedera}\ \emph {et~al.}(1987)\citenamefont {Bopegedera}, \citenamefont {Brazier},\ and\ \citenamefont {Bernath}}]{Bopegedera1987}%
  \BibitemOpen
  \bibfield  {author} {\bibinfo {author} {\bibfnamefont {A.~M.}\ \bibnamefont {Bopegedera}}, \bibinfo {author} {\bibfnamefont {C.~R.}\ \bibnamefont {Brazier}},\ and\ \bibinfo {author} {\bibfnamefont {P.~F.}\ \bibnamefont {Bernath}},\ }\href {https://doi.org/10.1021/j100295a026} {\bibfield  {journal} {\bibinfo  {journal} {J. Phys. Chem.}\ }\textbf {\bibinfo {volume} {91}},\ \bibinfo {pages} {2779} (\bibinfo {year} {1987})}\BibitemShut {NoStop}%
\bibitem [{\citenamefont {Brazier}\ and\ \citenamefont {Bernath}(2000)}]{Brazier2000}%
  \BibitemOpen
  \bibfield  {author} {\bibinfo {author} {\bibfnamefont {C.~R.}\ \bibnamefont {Brazier}}\ and\ \bibinfo {author} {\bibfnamefont {P.~F.}\ \bibnamefont {Bernath}},\ }\href {https://doi.org/10.1006/jmsp.2000.8076} {\bibfield  {journal} {\bibinfo  {journal} {J. Mol. Phys.}\ }\textbf {\bibinfo {volume} {201}},\ \bibinfo {pages} {116} (\bibinfo {year} {2000})}\BibitemShut {NoStop}%
\bibitem [{\citenamefont {Thompsen}\ \emph {et~al.}(2000)\citenamefont {Thompsen}, \citenamefont {Sheridan},\ and\ \citenamefont {Ziurys}}]{Thompsen2000}%
  \BibitemOpen
  \bibfield  {author} {\bibinfo {author} {\bibfnamefont {J.~M.}\ \bibnamefont {Thompsen}}, \bibinfo {author} {\bibfnamefont {P.~M.}\ \bibnamefont {Sheridan}},\ and\ \bibinfo {author} {\bibfnamefont {L.~M.}\ \bibnamefont {Ziurys}},\ }\href {https://doi.org/10.1016/S0009-2614(00)01113-1} {\bibfield  {journal} {\bibinfo  {journal} {Chem. Phys. Lett.}\ }\textbf {\bibinfo {volume} {330}},\ \bibinfo {pages} {373} (\bibinfo {year} {2000})}\BibitemShut {NoStop}%
\bibitem [{\citenamefont {Fernando}\ \emph {et~al.}(1991)\citenamefont {Fernando}, \citenamefont {Ram}, \citenamefont {O'Brien},\ and\ \citenamefont {Bernath}}]{Fernando1991}%
  \BibitemOpen
  \bibfield  {author} {\bibinfo {author} {\bibfnamefont {W.~T.}\ \bibnamefont {Fernando}}, \bibinfo {author} {\bibfnamefont {R.~S.}\ \bibnamefont {Ram}}, \bibinfo {author} {\bibfnamefont {L.~C.}\ \bibnamefont {O'Brien}},\ and\ \bibinfo {author} {\bibfnamefont {P.~F.}\ \bibnamefont {Bernath}},\ }\href {https://doi.org/10.1021/j100160a009} {\bibfield  {journal} {\bibinfo  {journal} {J. Phys. Chem.}\ }\textbf {\bibinfo {volume} {95}},\ \bibinfo {pages} {2665} (\bibinfo {year} {1991})}\BibitemShut {NoStop}%
\bibitem [{\citenamefont {Halfen}\ \emph {et~al.}(2001)\citenamefont {Halfen}, \citenamefont {Apponi}, \citenamefont {Thompsen},\ and\ \citenamefont {Ziurys}}]{Halfen2001}%
  \BibitemOpen
  \bibfield  {author} {\bibinfo {author} {\bibfnamefont {D.~T.}\ \bibnamefont {Halfen}}, \bibinfo {author} {\bibfnamefont {A.~J.}\ \bibnamefont {Apponi}}, \bibinfo {author} {\bibfnamefont {J.~M.}\ \bibnamefont {Thompsen}},\ and\ \bibinfo {author} {\bibfnamefont {L.~M.}\ \bibnamefont {Ziurys}},\ }\href {https://doi.org/10.1063/1.1419060} {\bibfield  {journal} {\bibinfo  {journal} {J. Chem. Phys.}\ }\textbf {\bibinfo {volume} {115}},\ \bibinfo {pages} {11131} (\bibinfo {year} {2001})}\BibitemShut {NoStop}%
\bibitem [{\citenamefont {Sheridan}\ \emph {et~al.}(2007)\citenamefont {Sheridan}, \citenamefont {Dick}, \citenamefont {Wang},\ and\ \citenamefont {Bernath}}]{Sheridan2007}%
  \BibitemOpen
  \bibfield  {author} {\bibinfo {author} {\bibfnamefont {P.~M.}\ \bibnamefont {Sheridan}}, \bibinfo {author} {\bibfnamefont {M.~J.}\ \bibnamefont {Dick}}, \bibinfo {author} {\bibfnamefont {J.~G.}\ \bibnamefont {Wang}},\ and\ \bibinfo {author} {\bibfnamefont {P.~F.}\ \bibnamefont {Bernath}},\ }\href {https://doi.org/10.1080/00268970701194418} {\bibfield  {journal} {\bibinfo  {journal} {Mol. Phys.}\ }\textbf {\bibinfo {volume} {105}},\ \bibinfo {pages} {569} (\bibinfo {year} {2007})}\BibitemShut {NoStop}%
\bibitem [{\citenamefont {Bunker}\ and\ \citenamefont {Jensen}(2006)}]{Bunker2006}%
  \BibitemOpen
  \bibfield  {author} {\bibinfo {author} {\bibfnamefont {P.~R.}\ \bibnamefont {Bunker}}\ and\ \bibinfo {author} {\bibfnamefont {P.}~\bibnamefont {Jensen}},\ }\href@noop {} {\emph {\bibinfo {title} {{Molecular Symmetry and Spectroscopy}}}},\ \bibinfo {edition} {2nd}\ ed.\ (\bibinfo  {publisher} {NRC Research Press},\ \bibinfo {year} {2006})\BibitemShut {NoStop}%
\bibitem [{\citenamefont {Baum}\ \emph {et~al.}(2021)\citenamefont {Baum}, \citenamefont {Vilas}, \citenamefont {Hallas}, \citenamefont {Augenbraun}, \citenamefont {Raval}, \citenamefont {Mitra},\ and\ \citenamefont {Doyle}}]{Baum2021}%
  \BibitemOpen
  \bibfield  {author} {\bibinfo {author} {\bibfnamefont {L.}~\bibnamefont {Baum}}, \bibinfo {author} {\bibfnamefont {N.~B.}\ \bibnamefont {Vilas}}, \bibinfo {author} {\bibfnamefont {C.}~\bibnamefont {Hallas}}, \bibinfo {author} {\bibfnamefont {B.~L.}\ \bibnamefont {Augenbraun}}, \bibinfo {author} {\bibfnamefont {S.}~\bibnamefont {Raval}}, \bibinfo {author} {\bibfnamefont {D.}~\bibnamefont {Mitra}},\ and\ \bibinfo {author} {\bibfnamefont {J.~M.}\ \bibnamefont {Doyle}},\ }\href {https://link.aps.org/doi/10.1103/PhysRevA.103.043111} {\bibfield  {journal} {\bibinfo  {journal} {Phys. Rev. A}\ }\textbf {\bibinfo {volume} {103}},\ \bibinfo {pages} {043111} (\bibinfo {year} {2021})}\BibitemShut {NoStop}%
\bibitem [{\citenamefont {Augenbraun}(2021)}]{Augenbraun2021a}%
  \BibitemOpen
  \bibfield  {author} {\bibinfo {author} {\bibfnamefont {B.~L.}\ \bibnamefont {Augenbraun}},\ }\emph {\bibinfo {title} {{Methods for Direct Laser Cooling of Polyatomic Molecules}}},\ \href {https://dash.harvard.edu/handle/1/37371102} {Ph.D. thesis},\ \bibinfo  {school} {Harvard University} (\bibinfo {year} {2021})\BibitemShut {NoStop}%
\bibitem [{\citenamefont {Ichino}\ \emph {et~al.}(2009)\citenamefont {Ichino}, \citenamefont {Gauss},\ and\ \citenamefont {Stanton}}]{Ichino09}%
  \BibitemOpen
  \bibfield  {author} {\bibinfo {author} {\bibfnamefont {T.}~\bibnamefont {Ichino}}, \bibinfo {author} {\bibfnamefont {J.}~\bibnamefont {Gauss}},\ and\ \bibinfo {author} {\bibfnamefont {J.~F.}\ \bibnamefont {Stanton}},\ }\href {https://doi.org/10.1063/1.3127246} {\bibfield  {journal} {\bibinfo  {journal} {J. Chem. Phys.}\ }\textbf {\bibinfo {volume} {130}},\ \bibinfo {pages} {174105} (\bibinfo {year} {2009})}\BibitemShut {NoStop}%
\bibitem [{\citenamefont {Vilas}\ \emph {et~al.}(2022)\citenamefont {Vilas}, \citenamefont {Hallas}, \citenamefont {Anderegg}, \citenamefont {Robichaud}, \citenamefont {Winnicki}, \citenamefont {Mitra},\ and\ \citenamefont {Doyle}}]{Vilas2022}%
  \BibitemOpen
  \bibfield  {author} {\bibinfo {author} {\bibfnamefont {N.~B.}\ \bibnamefont {Vilas}}, \bibinfo {author} {\bibfnamefont {C.}~\bibnamefont {Hallas}}, \bibinfo {author} {\bibfnamefont {L.}~\bibnamefont {Anderegg}}, \bibinfo {author} {\bibfnamefont {P.}~\bibnamefont {Robichaud}}, \bibinfo {author} {\bibfnamefont {A.}~\bibnamefont {Winnicki}}, \bibinfo {author} {\bibfnamefont {D.}~\bibnamefont {Mitra}},\ and\ \bibinfo {author} {\bibfnamefont {J.~M.}\ \bibnamefont {Doyle}},\ }\href {https://www.nature.com/articles/s41586-022-04620-5} {\bibfield  {journal} {\bibinfo  {journal} {Nature}\ }\textbf {\bibinfo {volume} {606}},\ \bibinfo {pages} {70} (\bibinfo {year} {2022})}\BibitemShut {NoStop}%
\bibitem [{\citenamefont {Augenbraun}\ \emph {et~al.}(2020{\natexlab{b}})\citenamefont {Augenbraun}, \citenamefont {Lasner}, \citenamefont {Frenett}, \citenamefont {Sawaoka}, \citenamefont {Miller}, \citenamefont {Steimle},\ and\ \citenamefont {Doyle}}]{Augenbraun2020Sisyphus}%
  \BibitemOpen
  \bibfield  {author} {\bibinfo {author} {\bibfnamefont {B.~L.}\ \bibnamefont {Augenbraun}}, \bibinfo {author} {\bibfnamefont {Z.~D.}\ \bibnamefont {Lasner}}, \bibinfo {author} {\bibfnamefont {A.}~\bibnamefont {Frenett}}, \bibinfo {author} {\bibfnamefont {H.}~\bibnamefont {Sawaoka}}, \bibinfo {author} {\bibfnamefont {C.}~\bibnamefont {Miller}}, \bibinfo {author} {\bibfnamefont {T.~C.}\ \bibnamefont {Steimle}},\ and\ \bibinfo {author} {\bibfnamefont {J.~M.}\ \bibnamefont {Doyle}},\ }\href {https://doi.org/10.1088/1367-2630/ab687b} {\bibfield  {journal} {\bibinfo  {journal} {New J. Phys.}\ }\textbf {\bibinfo {volume} {22}},\ \bibinfo {pages} {022003} (\bibinfo {year} {2020}{\natexlab{b}})}\BibitemShut {NoStop}%
\bibitem [{\citenamefont {Jadbabaie}\ \emph {et~al.}(2020)\citenamefont {Jadbabaie}, \citenamefont {Pilgram}, \citenamefont {K{\l}os}, \citenamefont {Kotochigova},\ and\ \citenamefont {Hutzler}}]{Jadbabaie2020}%
  \BibitemOpen
  \bibfield  {author} {\bibinfo {author} {\bibfnamefont {A.}~\bibnamefont {Jadbabaie}}, \bibinfo {author} {\bibfnamefont {N.~H.}\ \bibnamefont {Pilgram}}, \bibinfo {author} {\bibfnamefont {J.}~\bibnamefont {K{\l}os}}, \bibinfo {author} {\bibfnamefont {S.}~\bibnamefont {Kotochigova}},\ and\ \bibinfo {author} {\bibfnamefont {N.~R.}\ \bibnamefont {Hutzler}},\ }\href {https://iopscience.iop.org/article/10.1088/1367-2630/ab6eae/meta} {\bibfield  {journal} {\bibinfo  {journal} {New J. Phys.}\ }\textbf {\bibinfo {volume} {22}},\ \bibinfo {pages} {022002} (\bibinfo {year} {2020})}\BibitemShut {NoStop}%
\bibitem [{\citenamefont {Dick}\ \emph {et~al.}(2006)\citenamefont {Dick}, \citenamefont {Sheridan}, \citenamefont {Wang},\ and\ \citenamefont {Bernath}}]{Dick2006}%
  \BibitemOpen
  \bibfield  {author} {\bibinfo {author} {\bibfnamefont {M.~J.}\ \bibnamefont {Dick}}, \bibinfo {author} {\bibfnamefont {.~P.~M.}\ \bibnamefont {Sheridan}}, \bibinfo {author} {\bibfnamefont {J.-G.}\ \bibnamefont {Wang}},\ and\ \bibinfo {author} {\bibfnamefont {.~P.~F.}\ \bibnamefont {Bernath}},\ }\href {https://doi.org/10.1063/1.2189236} {\bibfield  {journal} {\bibinfo  {journal} {J. Chem. Phys}\ }\textbf {\bibinfo {volume} {124}},\ \bibinfo {pages} {174309} (\bibinfo {year} {2006})}\BibitemShut {NoStop}%
\end{thebibliography}%
\end{document}